\documentclass[conference]{IEEEtran}
\ifCLASSINFOpdf
\else
\fi
\usepackage{url}
\usepackage{amsmath}
\usepackage{algpseudocode}
\usepackage{algorithm}
\usepackage{hhline}
\usepackage{graphicx}
\graphicspath{ {figures/} }
\usepackage{svg}

\usepackage{amsmath,amssymb}
\usepackage{todonotes}
\usepackage{url}
\usepackage{amsmath}
\usepackage{algpseudocode}
\usepackage{algorithm}
\usepackage{hhline}
\usepackage{graphicx}
\graphicspath{ {figures/} }
\usepackage{svg}
\usepackage{subfig}
\usepackage{float}
\usepackage{adjustbox}
\usepackage{tabularx}
\usepackage{balance}

\usepackage{array,booktabs}
\newcolumntype{L}[1]{>{\raggedright\let\newline\\\arraybackslash\hspace{0pt}}m{#1}}
\newcolumntype{C}[1]{>{\centering\let\newline\\\arraybackslash\hspace{0pt}}m{#1}}
\newcolumntype{R}[1]{>{\raggedleft\let\newline\\\arraybackslash\hspace{0pt}}m{#1}}

\begin{document}
%
\title{\textbf{POWER-SUPPLaY:} Leaking Data from Air-Gapped Systems by Turning the Power-Supplies Into  Speakers}

%

\author{\IEEEauthorblockN{Mordechai Guri}
\IEEEauthorblockA{Ben-Gurion University of the Negev, Israel\\Cyber-Security Research Center\\
gurim@post.bgu.ac.il \\ Air-Gap research page: http://www.covertchannels.com \\ Demo video: http://www.covertchannels.com}}


%


\maketitle

\begin{abstract}
It is known that attackers can exfiltrate data from air-gapped computers through their speakers via sonic and ultrasonic waves. To eliminate the threat of such acoustic covert channels in sensitive systems, audio hardware can be disabled and the use of loudspeakers can be strictly forbidden. Such audio-less systems are considered to be \textit{audio-gapped}, and hence immune to acoustic covert channels.

In this paper, we introduce a technique that enable attackers leak data acoustically from air-gapped and audio-gapped systems. 
Our developed malware can exploit the computer power supply unit (PSU) to play sounds and use it as an out-of-band, secondary speaker with limited capabilities. The malicious code manipulates the internal \textit{switching frequency} of the power supply and hence controls the sound waveforms generated from its capacitors and transformers. Our technique enables producing audio tones in a frequency band of 0-24khz and playing audio streams (e.g., WAV) from a computer power supply without the need for audio hardware or speakers. Binary data (files, keylogging, encryption keys, etc.) can be modulated over the acoustic signals and sent to a nearby receiver (e.g., smartphone). We show that our technique works with various types of systems: PC workstations and servers, as well as embedded systems and IoT devices that have no audio hardware at all. We provide technical background and discuss implementation details such as signal generation and data modulation. We show that the POWER-SUPPLaY code can operate from an ordinary user-mode process and doesn't need any hardware access or special privileges. Our evaluation shows that using POWER-SUPPLaY, sensitive data can be exfiltrated from air-gapped and audio-gapped systems from a distance of five meters away at a maximal bit rates of 50 bit/sec.  
\end{abstract}


%
\IEEEpeerreviewmaketitle

	\section{Introduction}
Air-gapped computers are kept isolated from the Internet or other less secure networks. Such isolation is often enforced when sensitive or confidential data is involved, in order to reduce the risk of data leakage. Military networks such as the Joint Worldwide Intelligence Communications System (JWICS) \cite{Classifi75:online}, as well as networks within financial organizations, critical infrastructure, and commercial industries \cite{guri2017bridging,byres2013air}, are known to be air-gapped due to the sensitive data they handle. 

Despite the high degree of isolation, even air-gapped networks can be breached using complex attack vectors such as supply chain attacks, malicious insiders, and deceived insiders. Famous air-gap breaching cases include Stuxnet \cite{langner2011stuxnet} and Agent.BTZ \cite{grant2009cyber}, but other incidents have also been reported \cite{zaored,AFannyEq68:online}. In 2018, The US Department of Homeland Security accused Russian government hackers of penetrating America's power utilities \cite{Nobigdea65:online}. Due to reports in the Washington Post in November 2019, the Nuclear Power Corporation of India Limited (NPCIL) confirmed that the Kudankulam Nuclear Power Plant suffered a cyber-attack earlier that year \cite{AnIndian12:online}. 

\subsection{Air-Gap Covert Channels}
While the \textit{infiltration} of such networks has been shown to be feasible, the \textit{exfiltration} of data from non-networked computers or those without physical access is considered a challenging task. Over the years, different types of out-of-band covert channels have been proposed, exploring the feasibility of data exfiltration through an air-gap. \textit{Electromagnetic} methods that exploit electromagnetic radiation from different components of the computer are likely the oldest kind of air-gap covert channel researched \cite{guri2014airhopper,kuhn1998soft,kuhn2002compromising,vuagnoux2009compromising,guri2015gsmem}. Other types of \textit{optical} \cite{loughry2002information,Guri2017} and \textit{thermal} \cite{guri2015bitwhisper} out-of-band channels have also been studied. \textit{Acoustic} exfiltration of data using inaudible sound has also been explored in many studies \cite{hanspach2014covert,deshotels2014inaudible,madhavapeddy2005audio,Guri2018Mosquito}. The existing acoustic methods suggest transmitting data through the air-gap via high frequency sound waves generated by computer loudspeakers. Note that existing acoustic covert channels rely on the presence of audio hardware and loudspeakers in the compromised computer.

\subsection{Audio-Gap: Speaker-less, Audio-less Systems}
To cope with acoustic covert channels, common practices and security policies strictly prohibit the use of speakers on sensitive computers, in order to create a so-called 'audio-gapped' environment \cite{AirGapCo87:online,Guri:2018:BAM:3200906.3177230}. As an additional defensive measure, the audio chip may be disabled in the UEFI/BIOS to cope with the accidental attachment of loudspeakers to the \textit{line-out} connectors. Obviously, disabling the audio hardware and keeping speakers disconnected from computers can effectively mitigate the acoustic covert channels presented thus far \cite{Mindtheg67:online}.  

\subsection{Our Contribution}
In this paper, we introduce a new acoustic channel which doesn't require speakers or other audio related hardware. We show that malware running on a PC can exploit its power supply unit (PSU) and use it as an out-of-band speaker with limited capabilities. The malicious code intentionally manipulates the internal \textit{switching frequency} of the power supply and hence controls the waveform generated from its capacitors and transformers. This technique enables playing audio streams from a computer even when audio hardware is disabled and speakers are not present. We show that our technique works with various types of systems: PC workstations and servers, as well as embedded systems and IoT devices that have no audio hardware. Binary data can be modulated and transmitted out via the acoustic signals. The acoustic signals can then be intercepted by a nearby receiver (e.g., a smartphone), which demodulates and decodes the data and sends it to the attacker via the Internet.

The proposed method has the following unique characteristics:

\begin{itemize}
	\item {\textbf{Requires no audio hardware.}} The method allows malware to play audible and inaudible sounds from systems which are completely audio-gapped (e.g., speakers are disconnected) or systems that doesn't have any type of audio hardware (e.g., embedded devices). 
	
	\item {\textbf{Requires no special privileges.}} The method doesn't require special privileges or access to hardware resources. The transmitting code can be initiated from an ordinary user-space process and is highly evasive.
\end{itemize}

The rest of this paper is organized as follows: The attack model is discussed in Section \ref{sec:attack}. Related work is presented in Section \ref{sec:related}. Background on the power supply acoustics is provided in Section \ref{sec:tech}. Section \ref{sec:trans} and Section \ref{sec:rec}, respectively, contain details on The transmitter and receiver. Section \ref{sec:eval} describes the analysis and evaluation. Countermeasures are discussed in Section \ref{sec:counter}. We conclude in Section \ref{sec:conclusion}.

\section{Attack Model}
\label{sec:attack}
\begin{figure}
	\centering
	\includegraphics[width=\linewidth]{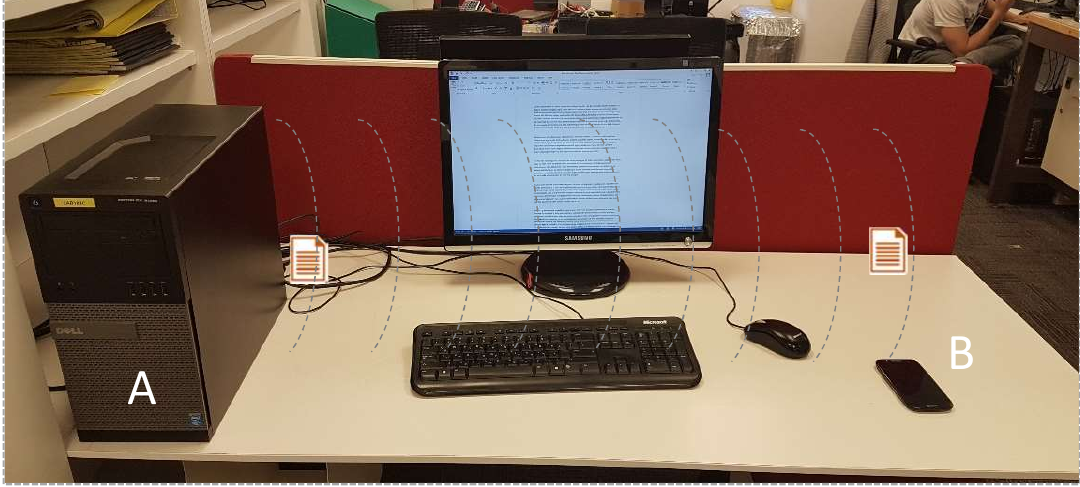}
	\caption{Exfiltration scenario: malware within the infected air-gapped, audio-gapped (speaker-less) computer (A) leaks a file through inaudible sound waves played through the power supply. The file is received by a nearby smartphone (B).}
	\label{fig:scenario}
\end{figure}

The capability of generating acoustic tones through power supplies can be considered a general contribution to the field of acoustic covert channels, regardless of its connection to the air-gap. However, in this paper, we investigate it as a method of exfiltrating information from air-gapped, audio-gapped systems. Similar to other covert communication channels, the adversarial attack model consists of a transmitter and a receiver. Typically in such scenarios, the transmitter is a computer, and the receiver is a nearby mobile phone belonging to an employee or visitor (Figure \ref{fig:scenario}). 

\subsubsection{Infection phase}
In a preliminary stage, the transmitter and receiver are compromised by the attacker. Infecting highly secure networks can be accomplished, as demonstrated by the attacks involving Stuxnet \cite {langner2011stuxnet} and Agent.Btz \cite{grant2009cyber}, and other attacks \cite{TheEpicT20:online,zaored,AFannyEq68:online}. In our case, the infected computer must be equipped with an internal power supply which exists in virtually every computerized system today. In addition, the mobile phones of employees are identified, possibly by social engineering techniques. The employees are assumed to carry their mobile phones around the workplace. These devices are then infected, either online, by exploiting a device's vulnerabilities, or by physical contact when possible. Infecting a mobile phone can be accomplished via different attack vectors, using emails, SMS/MMS, malicious apps, malicious websites, and so on \cite{provos2007ghost, cova2010detection, sood2011malvertising, peltier2006social, smutz2012malicious}.  
\subsubsection{Exfiltration phase} 
\begin{figure}
	\centering
	\includegraphics[width=1.0\linewidth]{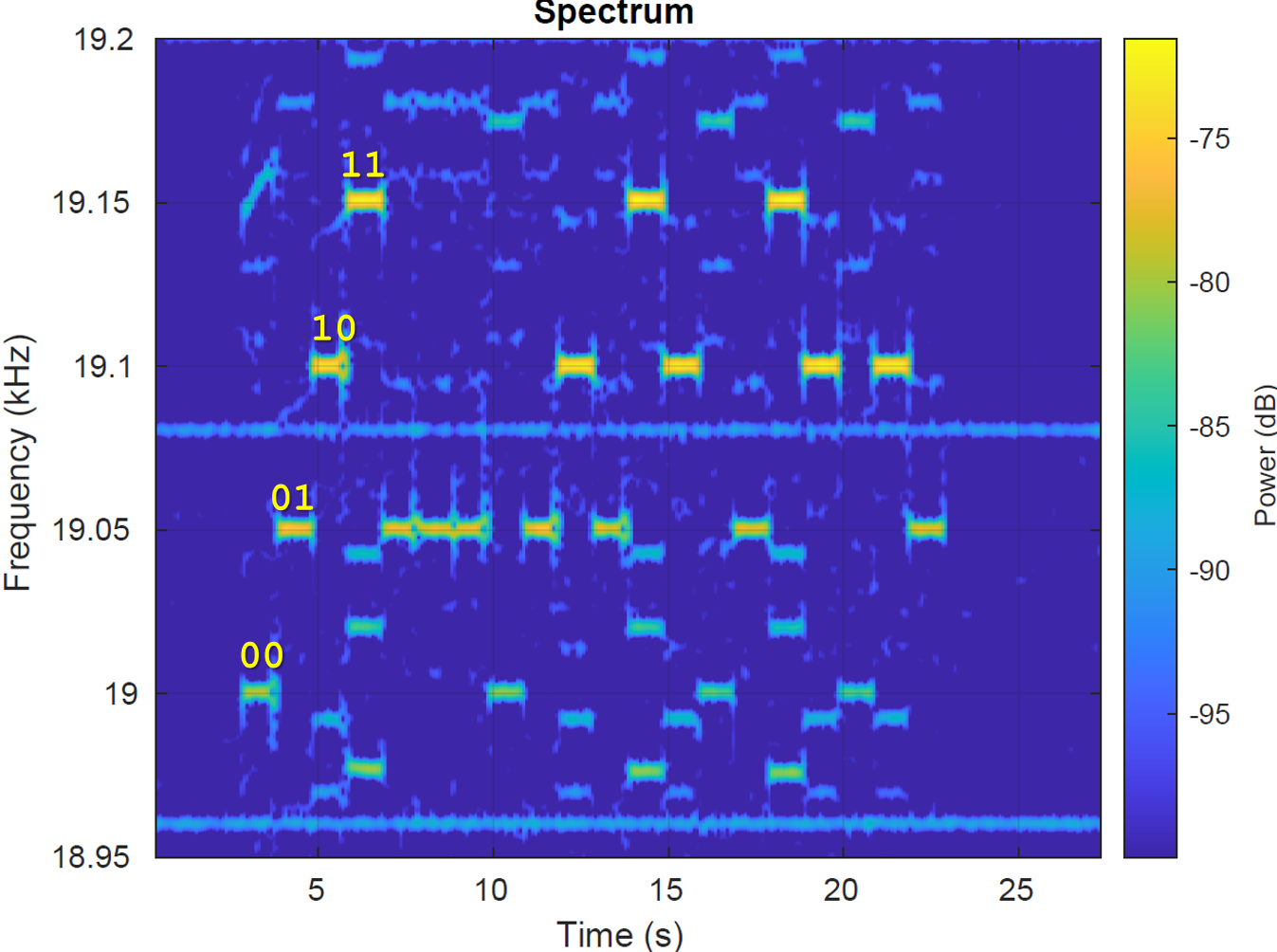}
	\caption{The information has been exfiltrated via covert ultrasonic sound signals played from the power supply. As can be seen in the spectrogram, four different frequencies are used for  modulation.}
	\label{fig:spec}
\end{figure}

In the exfiltration phase, the malware in the compromised computer gathers sensitive data of interest. The data can be files, keystroke logging, credentials (e.g., passwords), or encryption keys. The malware then modulates and transmits the data using the acoustic sound waves emitted from the computer's power supply (Figure \ref{fig:spec}). A nearby infected mobile phone detects the transmission, demodulates and decodes the data, and transfers it to the attacker via the Internet using mobile data or Wi-Fi. Note that in this paper, we demonstrate the attack model using a mobile phone receiver, a device which is commonly located in the vicinity of a computer. Other types of receivers are devices with internal or external microphones such as laptops and desktop workstations.

\begin{figure}
	\centering
	\includegraphics[width=1.0\linewidth]{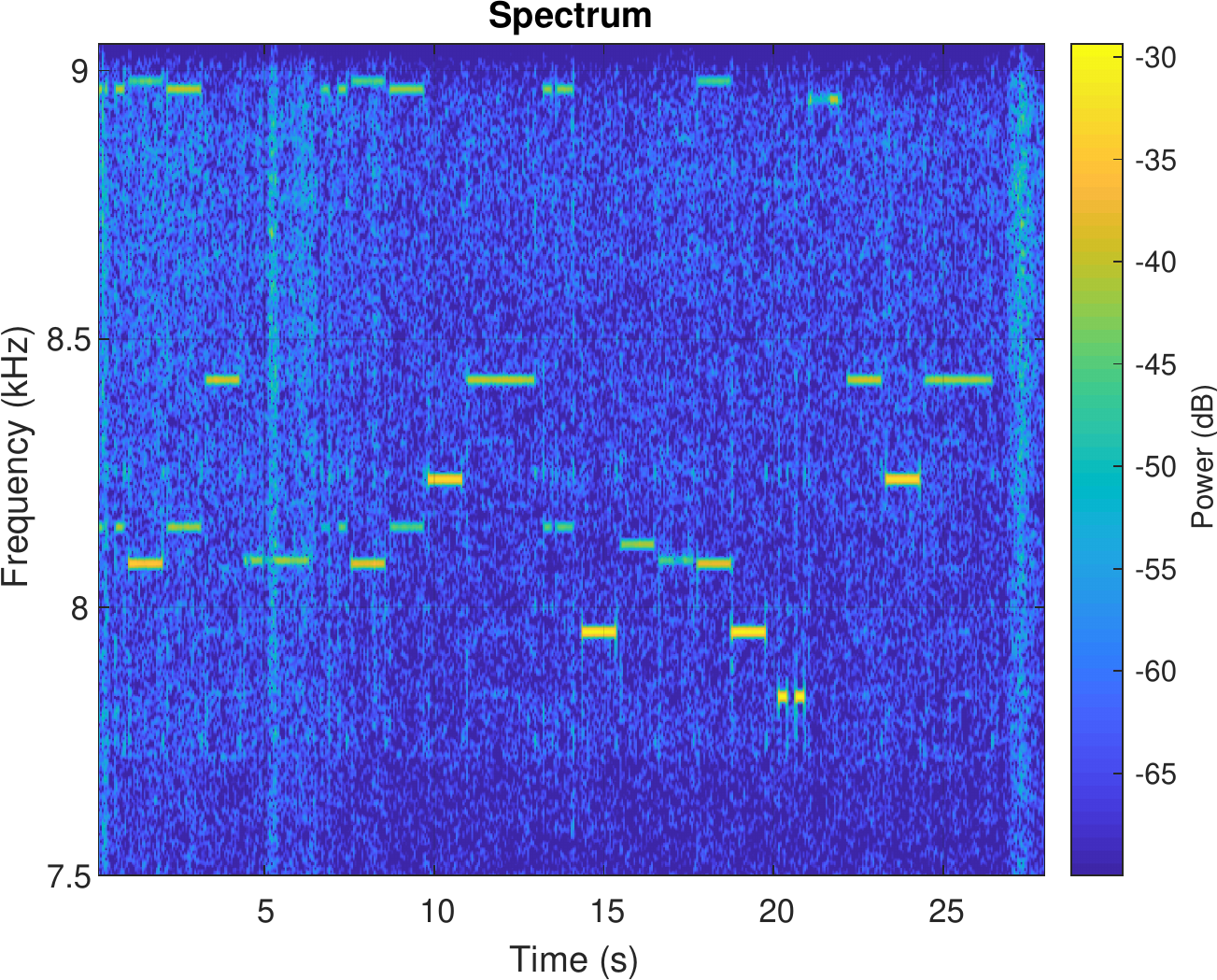}
	\caption{Part of the song 'Happy Birthday' as played through the power supply.}
	\label{fig:song}
\end{figure}

\section{Related Work}
\label{sec:related}
Covert channels in \textit{networked} environments have been widely discussed in professional literature for years \cite{giani2006data,murdoch2005embedding,zander2007survey}.  In these covert channels, attackers may hide data within existing network protocols (such as HTTPS, SMTP and DNS), conceal data within images (stenography), encode it in packet timings and so on.

Our work focuses on the challenge of leaking data from computers that have no network connectivity (air-gapped computers). Over the years different types of out-of-band covert channels have been proposed, allowing the attacker to bridge the air-gap isolation. The methods can be mainly categorized into electromagnetic and magnetic, optical, thermal, and acoustic covert channels.

Electromagnetic emissions are probably the oldest type of methods that have been explored academically with regard to air-gap communication. In a pioneer work in this field, Kuhn and Anderson \cite{kuhn1998soft} discuss hidden data transmission using electromagnetic emissions from video cards. Back in 2001, Thiele \cite{Tempestf48:online} utilized the computer monitor to transmit radio signals to a nearby AM radio receiver. AirHopper malware \cite{guri2014airhopper,guri2017bridging} introduced in 2014 by Guri et al, exploits video card emissions to bridge the air-gap between isolated computers and nearby mobile phones via FM radio signals. In a similar manner, GSMem \cite{guri2015gsmem}, Funthenna \cite{funtenna86:online} and USBee \cite{guri2016usbee}, introduce attack scenarios in which attackers use different sources of electromagnetic radiation on a computer's motherboard, as covert exfiltration channels. More recently, Guri et al proposed using the low-frequency magnetic fields emitted from the computer CPU for covert data exfiltration in order to leak data from Faraday caged air-gapped computers \cite{guri2019odini,guri2018magneto}. In 2018, Guri et al also introduced PowerHammer, a method in which malware on air-gapped computers exfiltrates data through the main power lines using conducted emissions \cite{guri2019powerhammer}. Several studies have proposed the use of optical emanation for air-gap communication. In 2002, Loughry and Umphress proposed the exfiltration of data by blinking the LEDs on the PC keyboard \cite{loughry2002information}. in 2019, Guri et al presented  CTRL-ALT-LED, a malware which can exfiltrate data from an air-gapped computer via the keyboard LEDs of modern USB keyboards \cite{guri2019ctrl}. In 2016, Shamir et al demonstrated how to establish a covert channel through the air-gap using a malware, remote lasers, and scanners \cite{Guri:2018:BAM:3200906.3177230}. Guri et al also presented covert channels that use the hard drive indicator LED \cite{Guri2017}, the router LEDs \cite{guri2018xled}, and security camera IR LEDs \cite{guri2019air} to leak data from air-gapped networks. VisiSploit \cite{guri2016optical} is another optical covert channel in which data is leaked through a hidden image projected on an LCD screen. Guri also showed how to exfiltrate data from air-gapped computers via fast blinking images \cite{guri2019optical}. BitWhisper \cite{guri2015bitwhisper} is a thermal based covert channel which enables covert communication between two adjacent air-gapped computers via the exchange of so-called 'thermal-pings.' 

\begin{table}[]
	\centering
	\caption{Summary of existing air-gap covert channels}
	\renewcommand{\arraystretch}{1.2}	
	\label{table-related}
	\begin{tabular}{l|l}
		\toprule
		Type               & Method  \\ \midrule[1pt]
		Electromagnetic    & \begin{tabular}[c]{@{}l@{}}AirHopper \cite{guri2014airhopper,guri2017bridging} (FM radio) \\ GSMem \cite{guri2015gsmem} (cellular frequencies)  \\ USBee \cite{guri2016usbee} (USB bus emission) \\ Funthenna \cite{funtenna86:online} (GPIO emission) \\
			PowerHammer (power lines) \cite{guri2019powerhammer} 
		\end{tabular}                                                                   \\ \midrule
		Magnetic           & \begin{tabular}[c]{@{}l@{}}MAGNETO \cite{guri2018magneto} (CPU-generated \\ magnetic fields)\\ ODINI \cite{guri2019odini} (Faraday shields bypass) \\ \end{tabular}            
		\\ \midrule
		Acoustic           & \begin{tabular}[c]{@{}l@{}} Ultrasonic (speaker-to-mic) \cite{hanspach2014covert,carrara2014acoustic}\\ MOSQUITO \cite{Guri2018Mosquito} (speaker-to-speaker) \\ Fansmitter \cite{guri2020fansmitter} (fans noise) \\ DiskFiltration \cite{guri2017acoustic} (hard disk noise) \\ \end{tabular} \\ \midrule
		Thermal            & BitWhisper  \cite{guri2015bitwhisper} (heat emission)                                                        \\ \midrule 
		Optical            & \begin{tabular}[c]{@{}l@{}}LED-it-GO \cite{Guri2017} (hard drive LED) \\ VisiSploit \cite{guri2016optical} (invisible pixels) \\ Fast blinking images \cite{guri2019optical} \\ Keyboard LEDs \cite{loughry2002information}\\  CTRL-ALT-LED: keyboard LEDs \cite{guri2019ctrl}\\  Router LEDs \cite{guri2018xled} \\                          
			aIR-Jumper \cite{guri2019air} (Infrared, security cameras)   \end{tabular} 
		\\ \midrule
			Vibrations (Seismic)           & \begin{tabular}[c]{@{}l@{}}AiR-ViBeR \cite{guri2020air} (Surface vibrations)

		    \end{tabular}                                                    \\ \bottomrule
	\end{tabular}
\end{table}

In acoustic covert channels, data is transmitted via audible or inaudible sound waves. In 2005, Madhavapeddy et al \cite{madhavapeddy2005audio} discuss 'audio networking,' which allows data transmission between a pair of desktop computers, using off-the-shelf speakers and a microphone. In 2013, Hanspach and Goetz \cite{hanspach2014covert} extended this method for near-ultrasonic covert networking between air-gapped laptops using built-in speakers and microphones. They created a mesh network and used it to implement an air-gapped key-logger which demonstrates the covert channel. The concept of communicating over inaudible sounds has been extended for different scenarios using laptops and smartphones \cite{deshotels2014inaudible}. In 2018, researchers presented MOSQUITO \cite{Guri2018Mosquito}, a covert communication channel between two air-gapped computers (without microphones) via so-called 'speaker-to-speaker' communication. In 2020, researcher presented a new type of vibrational (seismic) covert channel dubbed AiR-ViBeR. In this technique, data is  modulated in unnoticeable vibrations generated by the computer fans. The vibrations can be received and decoded by nearby smartphones via the integrated, sensitive accelerometers.   

\subsection{Speaker-less Methods}
All of the acoustic methods described above require the presence of external or internal speakers in the transmitting computer. This is considered a restrictive requirement, since loudspeakers are commonly forbidden in air-gapped computers \cite{Mindtheg67:online,AirGapCo87:online}. The elimination of loudspeakers is the most effective defense against the speaker-to-microphone and speaker-to-speaker covert channels discussed above \cite{Jumpingt83:online}.  

In 2016, Guri et al presented DiskFiltration, a method that uses the acoustic signals emitted from the hard disk drive (HDD) to exfiltrate data from air-gapped computers \cite{guri2017acoustic}. Although this method doesn't need a speaker, it is limited in terms of distance (up to two meters), and it doesn't work on newer technologies such as solid-state drives (SSDs). In 2016, Guri et al also introduced Fansmitter, malware which facilitates the exfiltration of data from an air-gapped computer via noise intentionally emitted from PC fans \cite{guri2020fansmitter}. In this method, the transmitting computer does not need to be equipped with audio hardware or an internal or external speaker. This method uses the internal fans, that exist in most laptops, desktops and server computers, and it is effective for longer distances of 6-7 meters and more. POWER-SUPPLaY differs from DiskFiltration and Fansmitter methods in two aspects. First, it uses only basic CPU instructions and does not use system resources like the HDD (DiskFiltration) or fans (Fansmitter). This makes it difficult for detection systems to identify the malicious activity of the transmitter. Second, POWER-SUPPLaY can operate at bit rates of 50 bit/sec which is much faster than Fansmitter (1 bit/sec) and DiskFiltration (3 bit/sec). In addition, POWER-SUPPLaY has the flexibility of generating acoustic tones in the 0-24 kHz frequency band, which implies that it is capable of generating both audible and inaudible sounds. Figure \ref{fig:song} shows the spectrogram of a part of the song 'Happy Birthday' as played form the power supply using the POWER-SUPPLaY technique.

Table 1 summarizes the existing covert channels for air-gapped computers.


\section{Technical Background}
\label{sec:tech}
In this section we provide the technical background on switch-mode power supplies (SMPSs), discuss the acoustic emission of SMPSs and describe the signal generation.
	
\subsection{Switch-Mode Power Supplies}
Computers consume power using their power supplies. Modern SMPS are used in all types of electronic equipment today, including computers, TVs, printers, Internet of Things (IoT) and embedded devices, and cell phone chargers. The advantages of SMPSs over the older linear power supplies include  their higher efficiency, smaller size, and lighter weight. An in-depth discussion on the design of SMPSs is beyond the scope of this paper, and we refer the interested reader to handbooks on this topic   \cite{handbookbillings2011switchmode}. Briefly, an SMPS transfers power from a 220V AC source to several DC loads while converting voltage and current characteristics. In the SMPS, the AC power passes through fuses and a line filter, and is then rectified by a full-wave bridge rectifier. The rectified voltage is applied to the power factor correction module and regulated via DC to DC converters.

\subsection{SMPS Acoustic Emission}
The DC supply from a rectifier or battery is fed to the inverter where it is turned on and off at high rates by the switching MOSFET or power transistors. This switching rate is also known as the \textit{switching frequency} of the SMPS. The typical switching frequency of a computer power supply is between 20 kHz and 20 MHz. 

The switching frequency affects, among others components in the SMPS, the transformers and capacitors. These are the primary sources of the acoustic noise generated by the SMPS.

\begin{figure}
	\centering
	\includegraphics[width=0.50\linewidth]{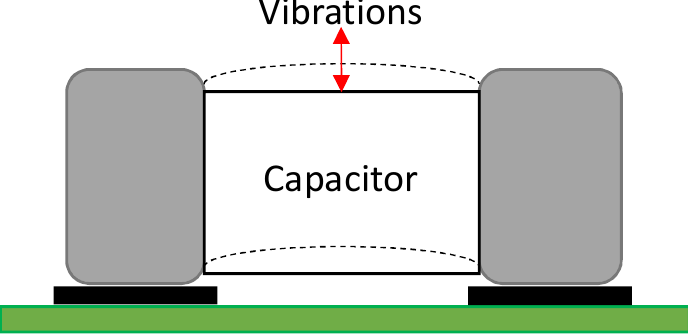}
	\caption{Illustration of the \textit{singing capacitor} phenomenon}
	\label{fig:dancing}
\end{figure}

\begin{itemize}
	\item {\textbf{Transformers' Audible Noise.}} 
	Transformers produce audible noise, since they contain many physically movable elements, such as coils, isolation tapes, and bobbins. The current in the coils, which occurs at the switching frequency, produces electromagnetic fields which generate repulsive and/or attractive forces between the coils. This can produce a mechanical vibration in the coils, ferrite cores, or isolation tapes.
	
	\item {\textbf{Capacitors Audible Noise.}} 
	Ceramic capacitors can produce audible noise, since they have Piezoelectric characteristics. The Piezoelectric acoustic effect on the capacitor is commonly described as "singing capacitors". 
	This noise is actually the result of vibrations of the capacitor on the Printed Circuit Board (PCB) that occur in normal working conditions. These vibrations causes capacitor
	displacement, as shown in Figure \ref{fig:dancing}. The frequency and amplitude of the displacement determine the acoustic waveform generated from the capacitors. When the vibration frequency occurs within the audible range, approximately 20 Hz – 20 kHz, it may also be heard as an audible hum. The range between 20 kHz and 24 kHz is considered as 'near-ultrasonic' and can not be heard by most humans.   
\end{itemize}

The typical switching frequency of SMPS during its normal operation is within the range of 20kHz-20MHz. Thus, the acoustic signal generated by SMPS is mainly in a frequency range of 20kHz and higher. This range is at the upper bound of human hearing and considered inaudible to adult humans.

\section{Transmission}
\label{sec:trans}

\subsection{Signal Generation}
Since modern CPUs are energy efficient, the momentary workload of the CPU directly affects the dynamic changes in its power consumption \cite{von2016variations}. By regulating the workload of the CPU, it is possible to govern its power consumption, and hence control the momentary \textit{switching frequency} of the SMPS. By intentionally starting and stopping the CPU workload, we are able to set the SMPS so it switches at a specified frequency and hence emit an acoustic signal and modulate binary data over it.

To generate a switching frequency $f_c$, we control the utilization of the CPU at a frequency correlated to $f_c$. To that end, $n$ worker threads are created where each thread is bound to a specific core. To generate the carrier wave, each worker thread overloads its core at a frequency $f_c$ repeatedly – alternating between applying a continuous workload on its core for a time period of $1/2f_c$ and putting its core in an idle state for a time period of $1/2f_c$. 

\begin{algorithm} 
	\caption{transmit (cores, freq, time)} 
	\label{alg0} 
	\begin{algorithmic}[1] 
		
		\State $ pthread\_barrier\_init(\&halfCycleBarrierHI, NULL, cores) $
		\State $ pthread\_barrier\_init(\&halfCycleBarrierLO, NULL, cores) $

		\For{$i \gets 1\ to\ cores$}
		\State { $ threadCreate(worker) $ }
		\EndFor
		
		\State $end = clock\_gettime() + time$
		\State $ halfCycleNano \gets 0.5*NANO\_PER\_SECOND/freq $
		\State $ cycleNano \gets NANO\_PER\_SECOND/freq $
		\While{$ clock\_gettime() < end $}
		\State $LO \gets 0$
		\State $pthread\_barrier\_wait(\&halfCycleBarrierLO)$
		\While{$ clock\_gettime()\%cycleNano < halfCycleNano $}
		\EndWhile

		\State $LO \gets 1$
		\State $pthread\_barrier\_wait(\&halfCycleBarrierHI)$
		\While{$ clock\_gettime()\%cycleNano >= halfCycleNano $}
		\EndWhile
		\EndWhile

	\end{algorithmic}
\end{algorithm}

\begin{algorithm} 
	\caption{worker  (threadState)} 
	\label{alg1} 
	\begin{algorithmic}[1] 
		
		\While {$ true $}
		\State $//sync\ threads\ on\ end\ of\ LO\ half\ cycle$
		\State $pthread\_barrier\_wait(\&halfCycleBarrierLO)$
		\State $ //HI\ half\ cycle\ -\ busy\ loop$
		\While {$ !LO $}
		\EndWhile 
		\State $//sync\ threads\ on\ end\ of\ HI\ half\ cycle$
		\State $pthread\_barrier\_wait(\&halfCycleBarrierHI)$ 
		\EndWhile

	\end{algorithmic}
\end{algorithm}

\begin{figure}
\centering
\includegraphics[width=0.68\linewidth]{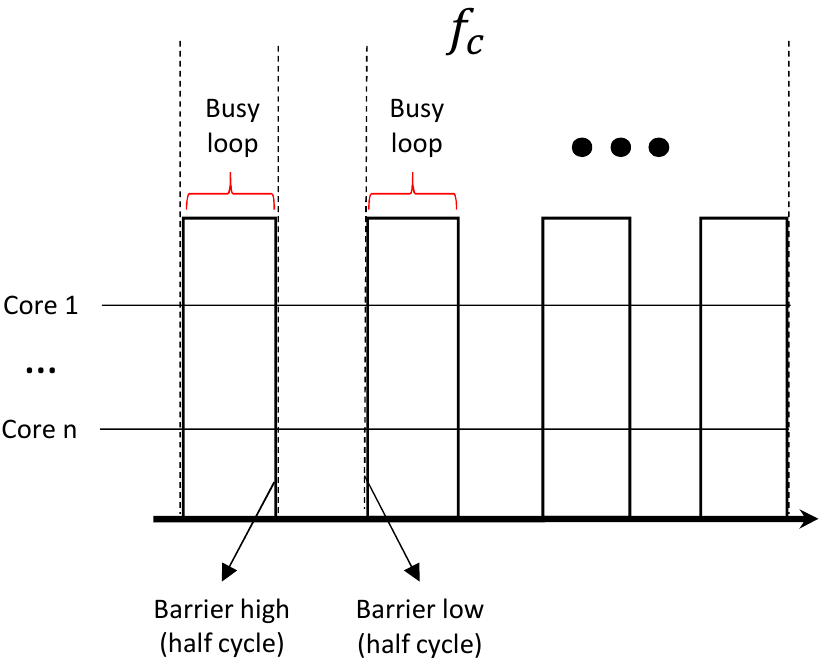}
\caption{$n$ cores utilization during a transmission}
\label{fig:utilization}
\end{figure}

Algorithms 1 shows the generation of acoustic tone ($freq$) for a time duration of ($time$) milliseconds, using ($nCores$) cores. At the beginning we initiate ($nCores$) threads and bound each thread to a specific core. The switching frequency is generated by each worker thread by employing busy loops and barrier objects. We overload the core using the busy waiting technique. This causes full utilization of the core for the time period and returns. The worker threads are synchronized with barrier objects, allowing them to start and stop the switching frequency altogether. The main thread governs the synchronization of the worker threads, changing the cycle state flag according to the barrier timing. This ensures that the generation of the duty cycles are precisely timed between the cores. Figure \ref{fig:utilization} illustrates the cores' utilization during the generation of a carrier wave $f_c$. Note that the power of the generated signal is correlated to the number of cores that participate in the transmission. The utilization of more cores yields more power consumption and hence more 'dancing capacitors'.   

Based on the algorithm above, we implemented a transmitter for Linux Ubuntu (version 16.04, 64 bit). We used the \texttt{sched\underline{ }setaffinity} system call to bind each thread to a CPU core. The affinity is the thread level attribute that is configured independently for each worker thread. To synchronize the initiation and termination of the worker threads, we used the thread barrier objects with \texttt{pthread$\_$barrier$\_$wait()} \cite{pthreadm53:online}. Note that the precision of \texttt{sleep()} is is not sufficient for our needs given the frequencies of the carrier waves which are at 24kHz or lower.

\subsection{Data Transfer}
\label{datatransfer}
The acoustic signal can be used to carry digital information using modulation schemes. For the data transfer we used the Frequency-Shift Keying (FSK) and the more advanced Orthogonal Frequency-Division Multiplexing (OFDM) modulation schemes.

\subsubsection{Frequency-Shift Keying}
In frequency-shift keying (FSK) the data is represented by a change in the frequency of a carrier wave. Recall that the transmitting code can determine the frequency of the generated signal. In FSK, each frequency represents a different symbol.

\begin{figure}
	\centering
	\includegraphics[width=0.8\linewidth]{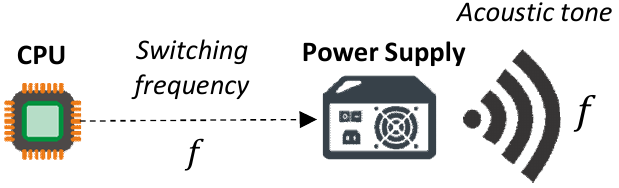}
	\caption{The signal generation}
	\label{fig:signalgneration}
\end{figure}

Figure \ref{fig:signalgneration} illustrates the generation of FSK signals. In this case, $n$ threads are bounded to the CPU cores. All threads are transmitting in the same carrier at given time, and hence generating a distinct switching frequency for each symbol. The generates acoustic tone is correlated with this switching frequency.

\begin{figure}
	\centering
	\includegraphics[width=\linewidth]{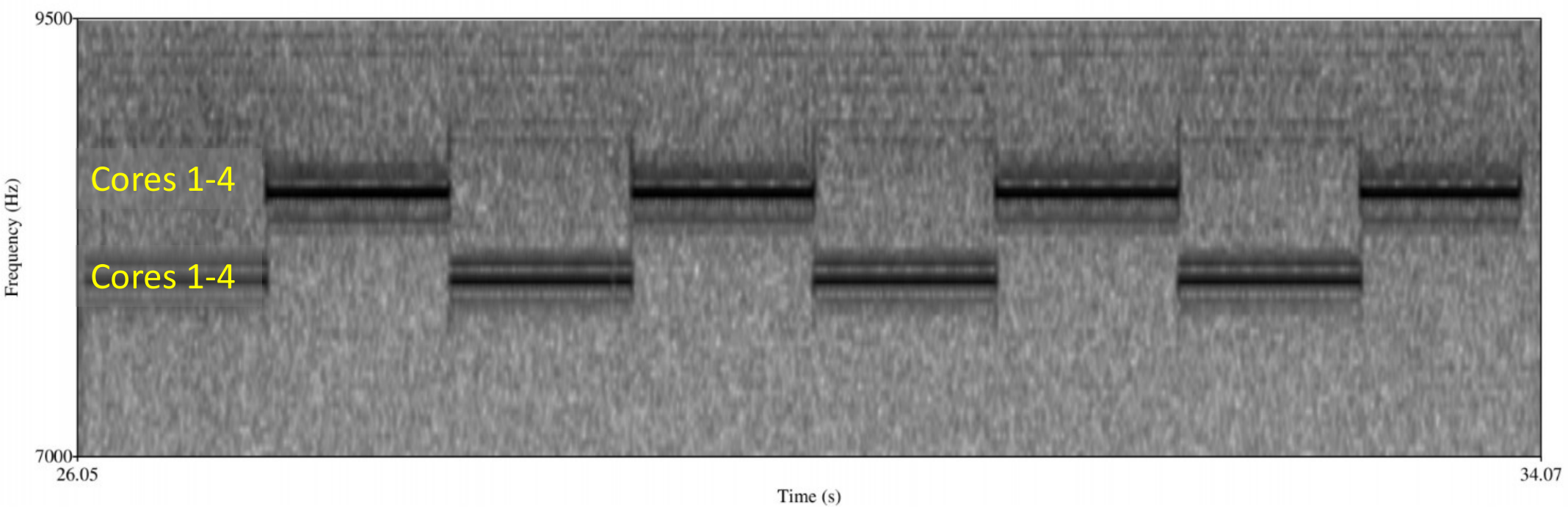}
	\caption{Spectrogram of the FSK modulation. The acousti signals generated from the PC power supply.}
	\label{fig:FSKSPEC}
\end{figure}

Figure \ref{fig:FSKSPEC} shows the time-frequency spectrogram of a binary sequence ('010101010') modulated with two frequency FSK (B-FSK) as transmitted from a PC with four cores. In this modulation, the frequencies 8500Hz and 8750Hz have been used to encode the symbols '0', '1', respectively.

%
%

%

\subsubsection{Orthogonal Frequency-Division Multiplexing}
We developed a fine-grained approach, in which we control the workload of each of the CPU cores independently from the other cores. Regulating the workload of each core separately enables greater control of the momentary switching frequency. By controlling the workload of each core separately, we can use a different sub-carrier for each transmitting core. This allows us to employ a more efficient modulation scheme such as orthogonal frequency-division multiplexing (OFDM).

In orthogonal frequency-division multiplexing data is represented by multiple carrier frequencies in parallel. In our case, we use different cores to transmit data in different sub-carriers. In each sub-carrier, we used on-off keying (OOK) to modulate the data. Note that since the sub-carriers' signals are generated in parallel, the maximal number of sub-carriers is equal to the number of cores available for the transmissions. 

\begin{figure}
	\centering
	\includegraphics[width=\linewidth]{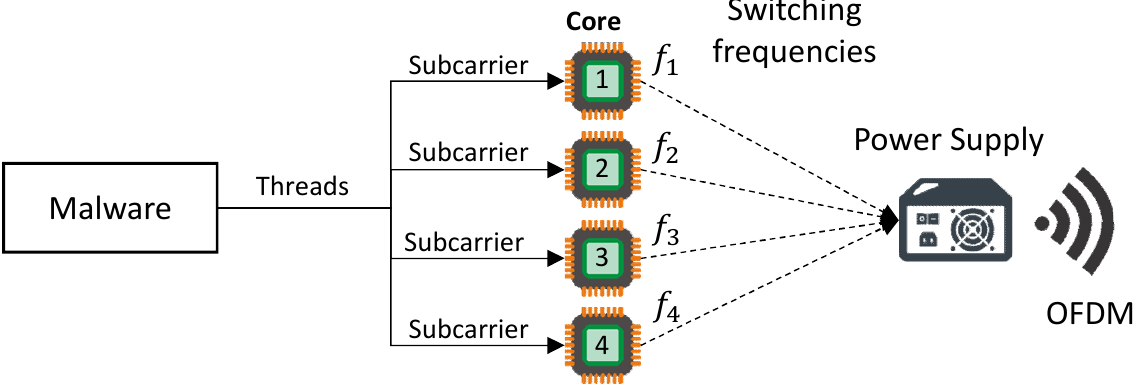}
	\caption{Illustration of the OFDM modulation}
	\label{fig:OFDM}
\end{figure}

Figure \ref{fig:OFDM} illustrates the generation of OFDM signals. In this case, $4$ threads are bounded to the  $4$ CPU cores. Each thread transmits in different sub-carrier, and hence generates different switching frequency. The acoustic tone is compound from all the switching frequencies generated by the sub-carriers.  

\begin{figure}
	\centering
	\includegraphics[width=0.8\linewidth]{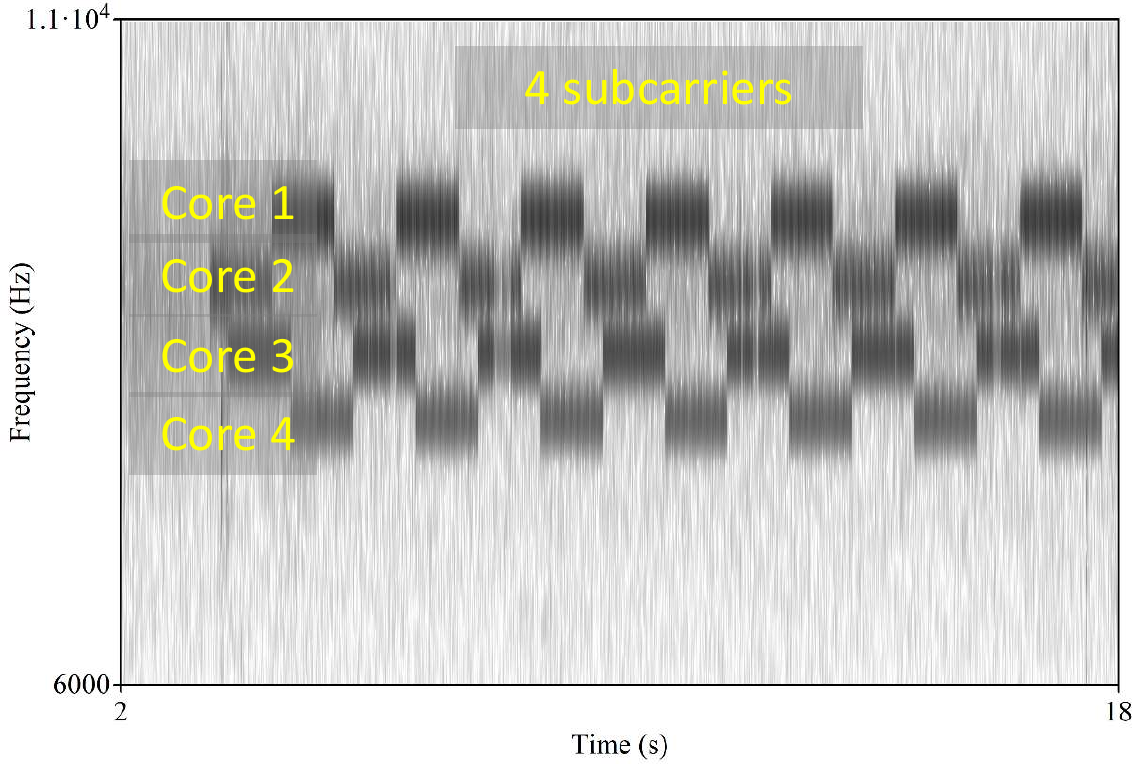}
	\caption{Spectrogram of the OFDM modulation}
	\label{fig:OFDMSPEC}
\end{figure}

Figure \ref{fig:OFDMSPEC} presents a binary sequence  modulated with OFDM with four sub-carriers as transmitted from a PC with four cores. In this modulation, 8000Hz (core 1), 8200Hz (core 2), 8400Hz (core 3) and 8600Hz (core 4) have been used to encode the symbols '00', '01', '10' and '11', respectively. 

\subsection{Transmission Protocol}
We transmit the data in small packets composed of a preamble, a payload, and a cyclic redundancy check (CRC) code.  

\begin{figure}
	\centering
	\includegraphics[width=\linewidth]{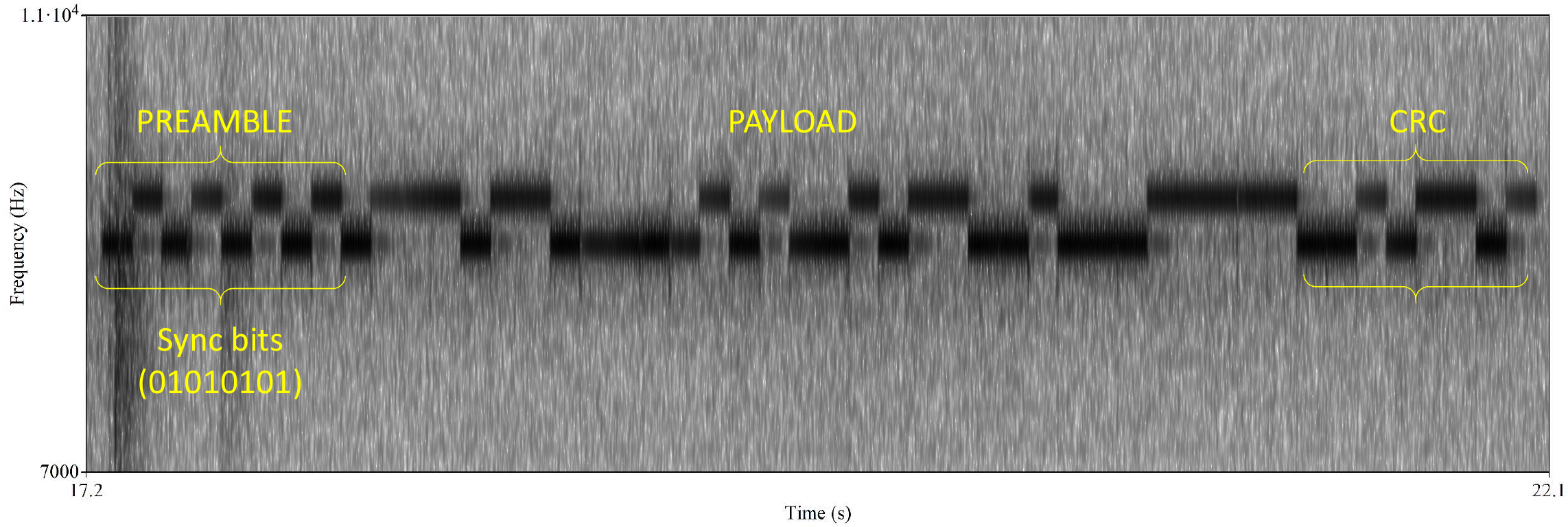}
	\caption{Spectrogram of a frame modulated with B-FSK as transmitted from a PC power supply}
	\label{fig:frame}
\end{figure}

\begin{itemize}
	\item	\textbf{Preamble.} A preamble header is transmitted at the beginning of every packet. It consists of a sequence of eight alternating bits ('10101010') which helps the receiver determine the carrier wave frequency and amplitude. In addition, the preamble allows the receiver to detect the beginning of a transmission. 
	Note that in our covert channel the amplitude of the carrier wave is unknown to the receiver in advance, and it mainly depends on what type of transmitting computer is used, the number of cores participating in the transmission, and the distance between the transmitter and the receiver. These parameters are synchronized with the receiver during the preamble.  
	\item \textbf{Payload.} The payload is the raw data to be transmitted. In our case, we arbitrarily choose 32 bits as the payload size. 
	\item \textbf{CRC.} For error detection, an eight bit CRC (cyclic redundancy check) is added to the end of the frame. The receiver calculates the CRC for the received payload, and if it differs from the received CRC bit, an error is detected. 
	A spectrogram of a full frame transmitted from a computer is presented in Figure \ref{fig:frame}. In this case,  a B-FSK was used with frequencies of 8400Hz and 8600Hz.   
	
\end{itemize}

\begin{figure}
	\centering
	\includegraphics[width=\linewidth]{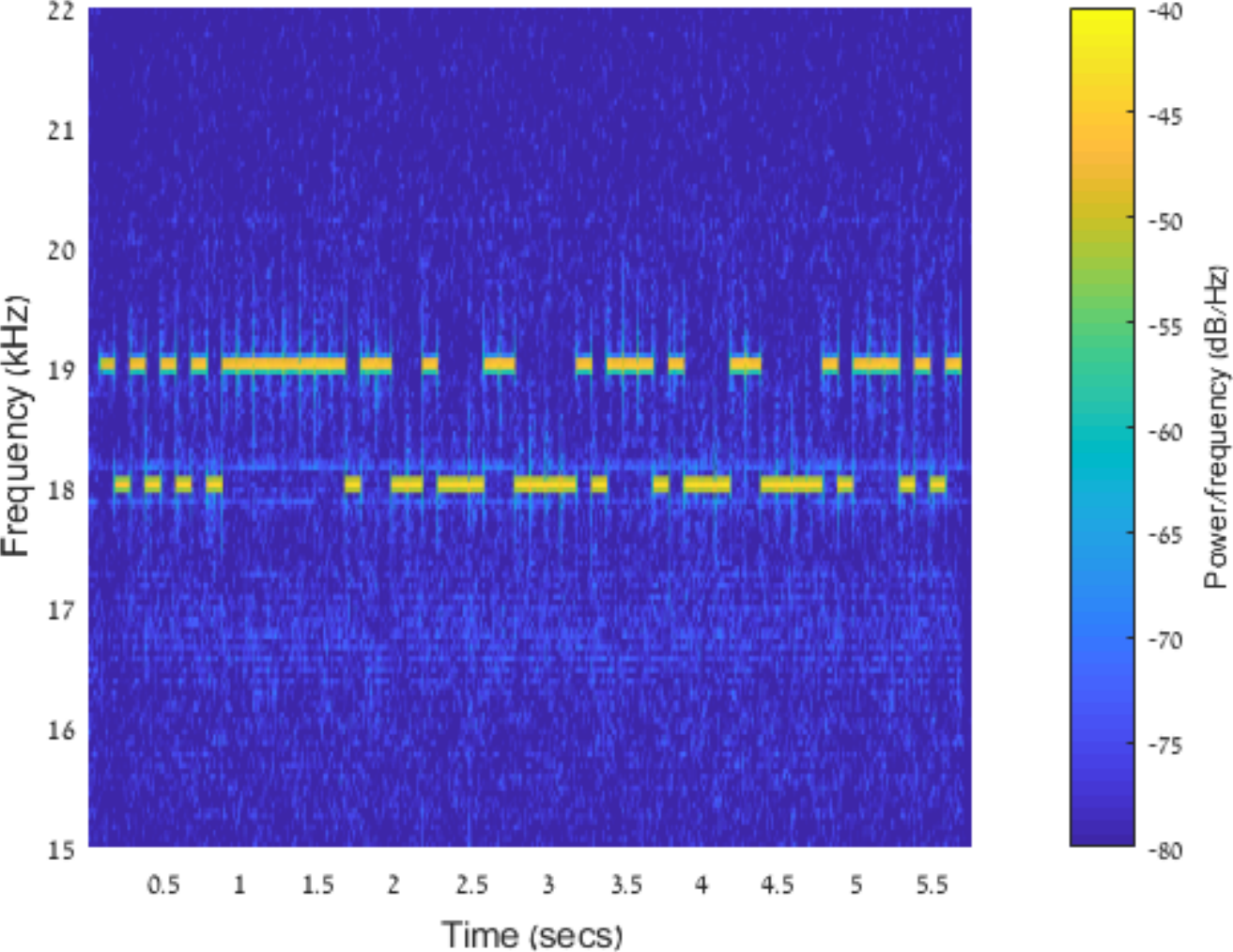}
	\caption{Spectrogram of a frame modulated with B-FSK as transmitted from a PC power supply}
	\label{fig:frame2}
\end{figure}

\section{Audio Formats}
We implemented a program which plays digital audio streams through the power supply. Our implementation supports the Waveform Audio File (WAV) format. The most common WAV format encodes uncompressed audio in pulse code modulation (PCM). There are two fundamental parameters which should be considered when playing audio streams: the \textit{sample rate} and the \textit{bit depth}. We briefly discuss each with regard to playing audio through the power supply.  

\subsection{Sample Rate}
The sampling rate (measured in kHz) of an audio file is the number of samples of audio carried per second. According to the Nyquist Theorem, the maximum frequency that can be represented in an audio file is half of the sample rate. Most sound cards can play audio streams at 48kHz, which is the sample rate used for DVDs. In the case of SMPS, the maximum playing frequency depends on the maximum rate at which we can start and stop utilizing the CPU cores, which directly affects the SMPS switching frequency. Using the barriers and busy loop technique described above, we could switch each CPU core at 100kHz, and hence are capable of playing audio streams at lower rates of 48kHz.

\subsection{Bit Depth}
In digital audio the bit-depth represents the number of possible amplitude values for a sample. The amplitude of each sample is encoded by the number of bits, which is the bit depth of the audio stream. In standard sound cards, the amplitude of a sample is sent to the digital-to-analog converter (DAC)  component and played through the loudspeakers. The loudspeakers' membranes vibrate in a power correlated to the amplitude that produces the desired signal. In the case of SMPS, the amplitude of the signal generated from the capacitors is constant and produces square acoustic waves. This represents a simple audio stream with a depth of only 1-bit (signal/no-signal). In order to play complex audio streams with a wider bit depth, we implemented two different modulation techniques; (1) amplitude modulation (AM), and (2) pulse width modulation (PWM).

\subsubsection{Amplitude Modulation}
\begin{figure}[!h]
	\centering
	\includegraphics[width=0.75\linewidth]{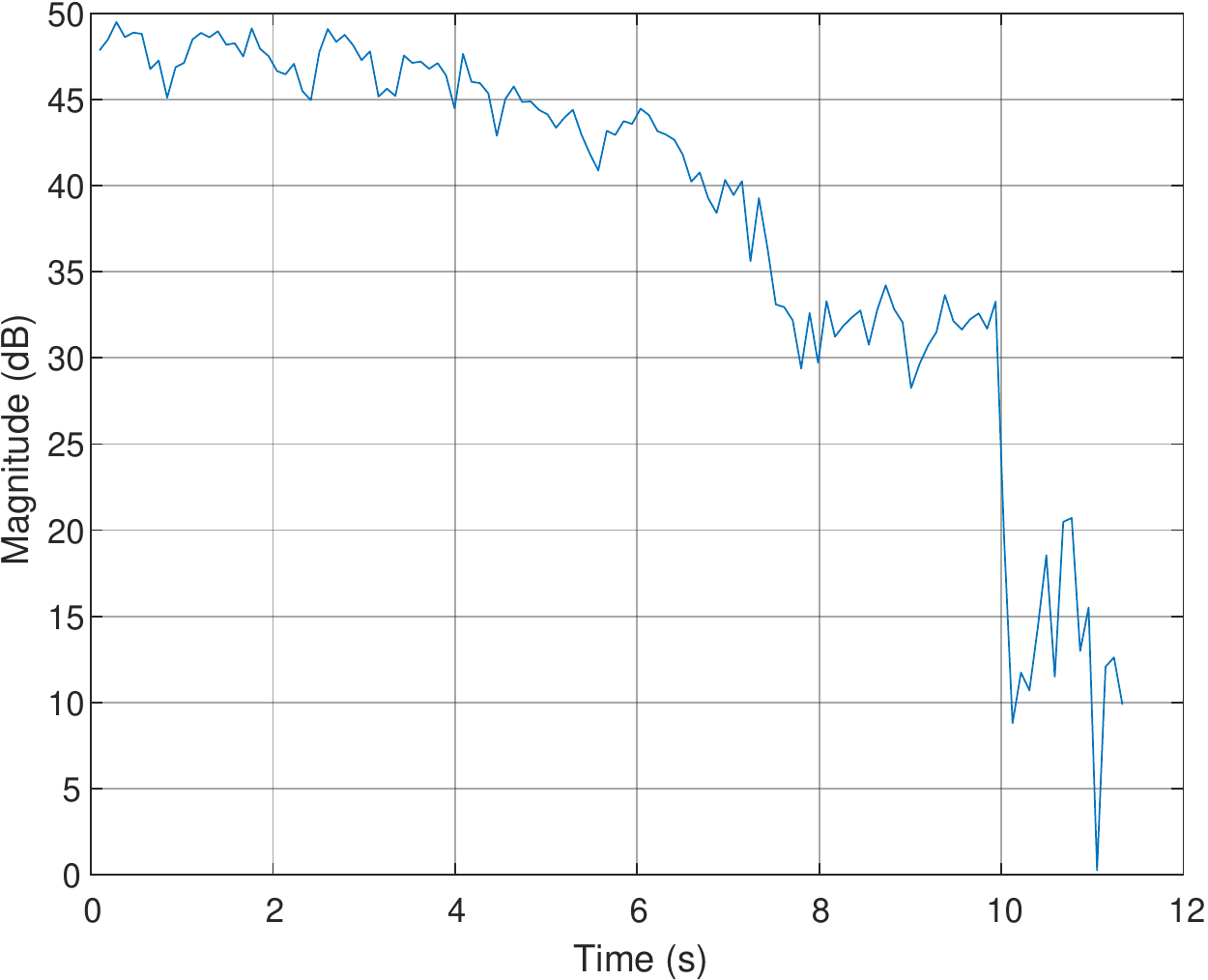}
	\caption{Three amplitudes generated with the AM technique}
	\label{fig:AM}
\end{figure}

As described in the signal generation section, the number of cores used for  signal generation affects the strength of the signal. This is because the more cores that are utilized, the more capacitors and transformers that are switched in the SMPS. This enables some control of the \textit{amplitude} generated for each sample. Given $N$ virtual cores, it is possible to play each sample at $N$ different amplitudes and hence play audio streams with a bit depth of roughly $N$. Figure \ref{fig:AM} shows the waveform of a signal with three amplitudes generated with the AM technique. In this case, the signal is generated using four, two and one cores. 

\subsubsection{Pulse Width Modulation}
PWM is a method of reducing the average power delivered by an electrical signal, by effectively breaking it into discrete parts. The average value of voltage (and current) fed to the load is controlled by turning the switch between the supply and load on and off at a fast rate. The longer the switch is on compared to the time it is off, the greater the total amount of power supplied to the load.

By carefully timing a short pulse (i.e., going from one output level to the other and back), the end result corresponds to intermediate sound levels, functioning as a crude DAC. PWM allows an approximate playback of WAV and PCM audio.

\begin{figure}
	\centering
	\includegraphics[width=0.65\linewidth]{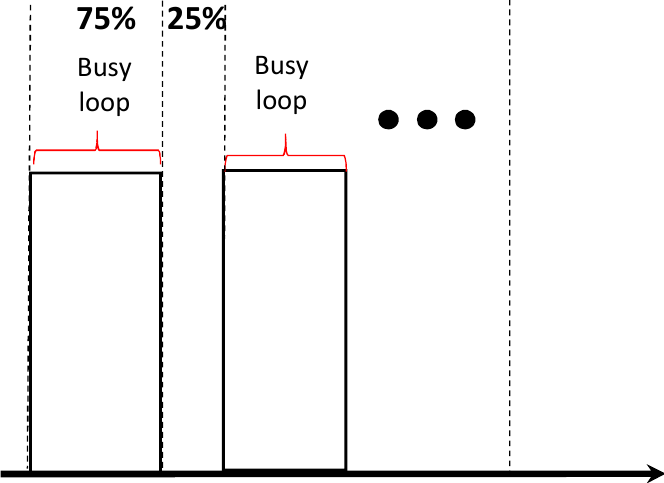}
	\caption{Illustration of a PWM modulated signal}
	\label{fig:PWMpic}
\end{figure}

\begin{figure}[!h]
	\centering
	\includegraphics[width=\linewidth]{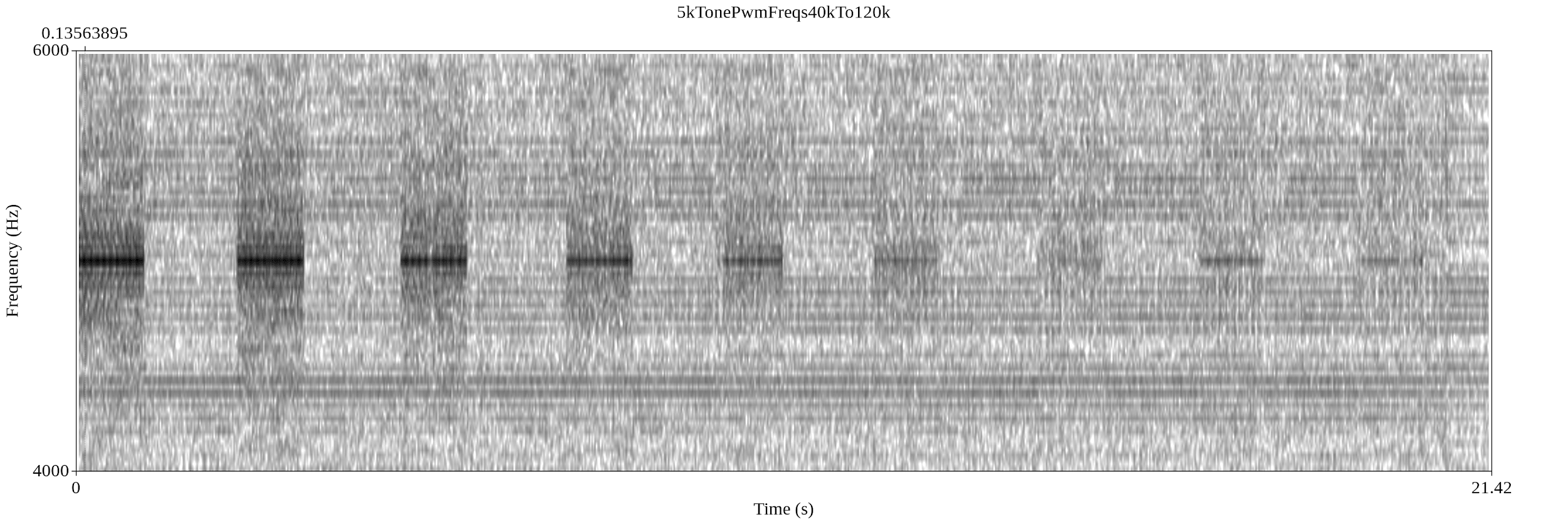}
	\caption{Eight levels of sound using the PWM technique}
	\label{fig:PWM}
\end{figure}

The sound level of a sample is determined by the duration of the CPU utilization. This is the \textit{duty cycle} of the signal. A short duty cycle corresponds to low power (low volume), because the power is off for most of the time, while a long duty cycle corresponds to high power (high volume). We calculate the duty cycle required for each sample given its amplitude value. Figure \ref{fig:PWMpic} illustrates a signal generated with a duty cycle of 75\% which implies a signal which is 25\% stronger than an average signal in the stream. Figure \ref{fig:PWM} contains the spectrogram of a signal with eight decreasing amplitudes, generated with the PWM technique. In this case, the signal is generated using eight duty cycles which decrease over time.  

\subsection{Play WAV via Power-Supply}
Based on the AM and PWM techniques described above, we implemented a program which plays PCM and WAV files through the power supply. Our \texttt{PowerSupplay [AM | PWM] wav\_file.wav} program determines which type of modulation to use and plays the \textit{wav\_file} using the corresponding technique.

\section{Reception}
\label{sec:rec}
We implemented a receiver as an app for the Android OS. The main functionality of the receiver is (1) sampling the signal, (2) performing signal processing, (3) detecting the preamble header, (4) demodulating the payload, and (5) handling errors.  

The main functionality of the receiver is run in a separate thread. It is responsible for data sampling, signal processing, and data demodulation. An outline of the receiver is presented in Algorithm \ref{alg1}; this is followed by a description. 

\begin{algorithm} 
	\renewcommand\thealgorithm{1}
	\caption{}
	\label{alg1} 
	\begin{algorithmic}[1] 
		\Procedure{Receiver}{}
		\State {$raw\_signal = sample(44.1kHz)$}

		\State {$signal = FFT(raw\_signal)$}
		
		\If {$(state == PREAMBLE)$}
		\If {$(DetectPreamble(signal) == true)$}
		\State {$T, f_{0}, f_{1} = ExtractChannelParams(signal)$}
		\State {$SetState(DEMODULATE)$}
		\EndIf
		\EndIf

		\If {$ (state == DEMODULATE)$}
		\State{$bit = Demodulate(signal)$}
		\State{$data.add(bit)$}
		\If {$(data.size\%(32 + 8)==0)$}
		\State{$SetState(PREAMBLE)$}
		\EndIf
		\If {$(SignalLost(signal)$}
		\State{$SetState(PREAMBLE)$}
		\EndIf
		
		\EndIf

		\EndProcedure
	\end{algorithmic}
\end{algorithm}

\subsubsection{Signal Sampling and Downsampling}
In order to analyze the waveforms in the frequency domain, we record data at a rate of 44.1kHz and save it to a temporary buffer.

\subsubsection{Signal Processing} The first step is to measure the signal and transfer it to the frequency domain. This step includes performing a fast Fourier transform (FFT) on the sampled signal. The signal measured is stored in a buffer after applying a filter for noise
mitigation. This data is used later in the demodulation routines. The noise mitigation function is applied to the current sample by averaging it with the last $w$ samples.

\subsubsection{Preamble Detection} In the $PREAMBLE$ state the receiver
searches for a preamble sequence to identify a frame header
(lines 5-8). If the preamble sequence '10101010' is detected, the
state is changed to $DEMODULATE$ to initiate the
demodulation process. Based on the preamble sequence, the receiver determines the channel parameters, signal time, ($T$) and frequencies $f_{0}$ and $f_{1}$.
\subsubsection{Demodulation} In the $DEMODULATE$ state the payload is
demodulated given the signal parameters retrieved in the
$PREAMBLE$ state. The demodulated bit is added to the current payload. When the payload of 32 bits and the 8-bit CRC are received, the algorithm returns back to the preamble detection state ($PREAMBLE$).
\subsubsection{Error Handling}
The $SignalLost()$ function returns true if, during the data
reception, the signal power measured in the carrier frequency is weaker than the amplitude of the '0's from the preamble for three seconds
straight. In this case, any partially received data is discarded, and the function returns to the
$PREAMBLE$ state. Signal loss may occur if the malware stops
the transmission (e.g., for stealth or due to the computer
shutting down). Signal loss may also occur if the receiving smartphone
has been moved away from the transmitting computer.

\begin{figure}
	\centering
	\includegraphics[width=\linewidth]{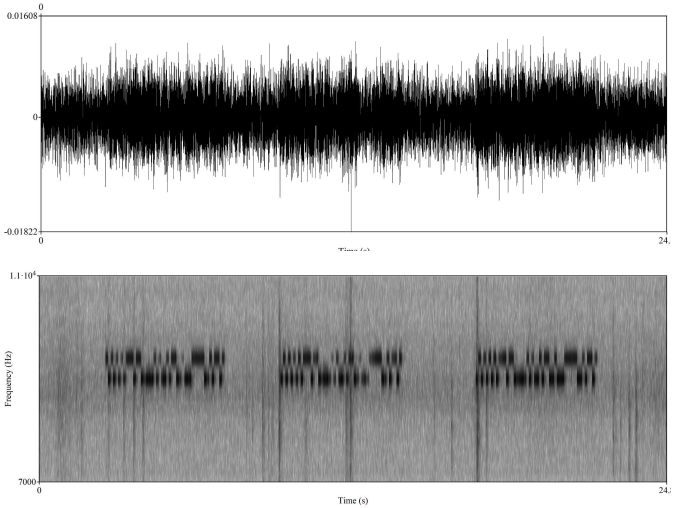}
	\caption{The raw signal and the demodulated bits}
	\label{fig:ba}
\end{figure}

The demodulation is depicted in Figure \ref{fig:ba}. The upper image presents the raw acoustic signal recorded by a nearby smartphone. The spectrogram in the lower image shows the demodulated bits.

\section{Evaluation}
\label{sec:eval}
In this section we provide an analysis and evaluation of the acoustic waveforms generated through the power supply. We also present the experimental setup and discuss the measurements and results.

\subsection{Measurements Setup}
\label{sec:measurement_setup}

\subsubsection{Transmitters}
We examined the acoustic signal generated from six types of computers: three standard desktop workstations (PC-1, PC-2, and PC-3), a server machine with multi-core processors (Server), a small form factor computer (NUK), and a low power embedded device (IOT). A detailed list of the computers tested and their technical specifications is provided in Table \ref{table:list}.

\begin{table*}[]
	\centering
	\caption{The computers used in the experiments}
	\label{table:list}
	\begin{tabular}{@{}lllll@{}}
		\toprule
		\#     & Type              & Model              & PSU                    & CPU                                                                                                       \\ \midrule
		PC-1   & Desktop PC        & Silverstone        & FSP300-50HMN 300W      & \begin{tabular}[c]{@{}l@{}}Intel Core i7-4770 CPU@ 3.4GHz (4 cores) \end{tabular}              \\
		PC-2   & Desktop PC        & Infinity           & PK-450 450W                & 
		Intel Core i7-4770 CPU 3.4GHz (4 cores)                                                                                                    \\
		PC-3   & Desktop PC        & Lenovo             & ATX POWER SUPPLY 280 WATT                 & Intel Core Quad-Q9550 CPU @2.83GHz (4 cores)                                                                                                    \\
		Server & Server            & IBM                & DPS-750AB A 750Wx2     & \begin{tabular}[c]{@{}l@{}}Intel Xeon CPUE5-2620 12 cores (24 threads)\end{tabular}                     \\
		NUK    & Small form factor & Lenovo ThinkCentre & PA-1650-72 20V 3.25A                 & Intel Core i7 -4785T (4 cores)                                                                                                    \\
		IOT    & Embedded/IoT      & Raspberry Pi 3     & DSA-13 PFC-05, 5V 2.5A & \begin{tabular}[c]{@{}l@{}}Quad Core Broadcom BCM2837, 64-bit, ARMv8, processor Cortex A53\end{tabular} \\ \bottomrule
	\end{tabular}
\end{table*}

\subsubsection{Receiver}
During the tests, the acoustic signals were recorded with a Samsung Galaxy S7 (G930F) smartphone, running the Android OS. The audio sampling was performed using the smartphone's internal microphone at a sampling rate of 44.1kHz. We used the \textit{MathWorks MATLAB} environment for signal processing, and the \textit{Praat} framework for additional spectral analysis \cite{Praatdoi67:online}.  

\subsection{Spectral View}
Power supplies are not intended to generate sound and hence are limited in terms of the acoustic signal they can produce. Unlike loudspeakers which can generate sound waves in the entire spectrum of 0-24kHz, a good quality signal generated from the PSU transformers and capacitors is only obtained at certain fragmented bands of the spectrum. These ranges are largely specific to each PSU, and dependent on their internal design, structure, and the type of transformers and capacitors.

\begin{figure}[]
	\centering
		\includegraphics[width=\linewidth]{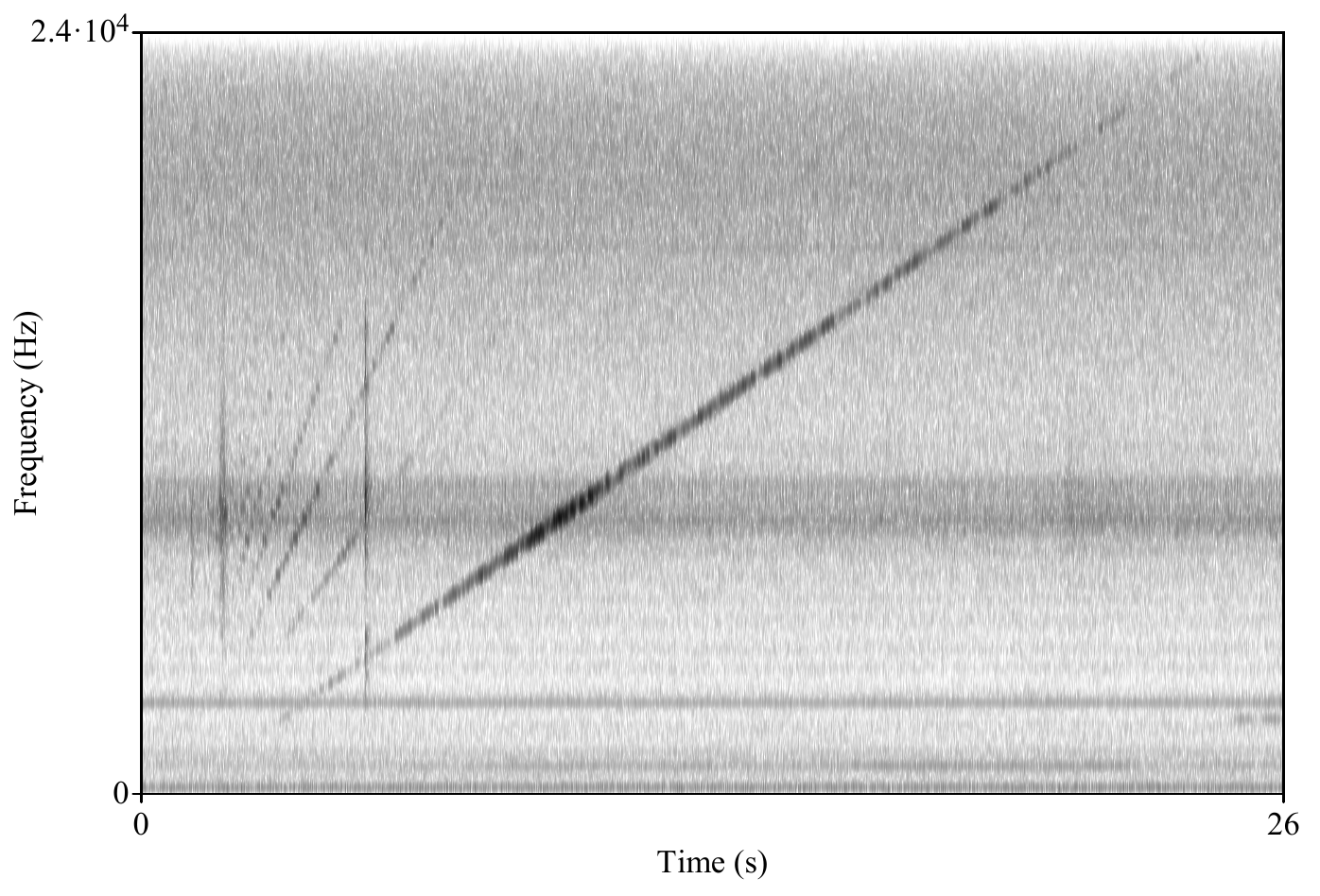}
		\caption{The sweep signal recorded from PC-1 (spectrogram view)}
	\label{fig:as1}
\end{figure}

\begin{figure}[]
	\centering
		\includegraphics[width=\linewidth]{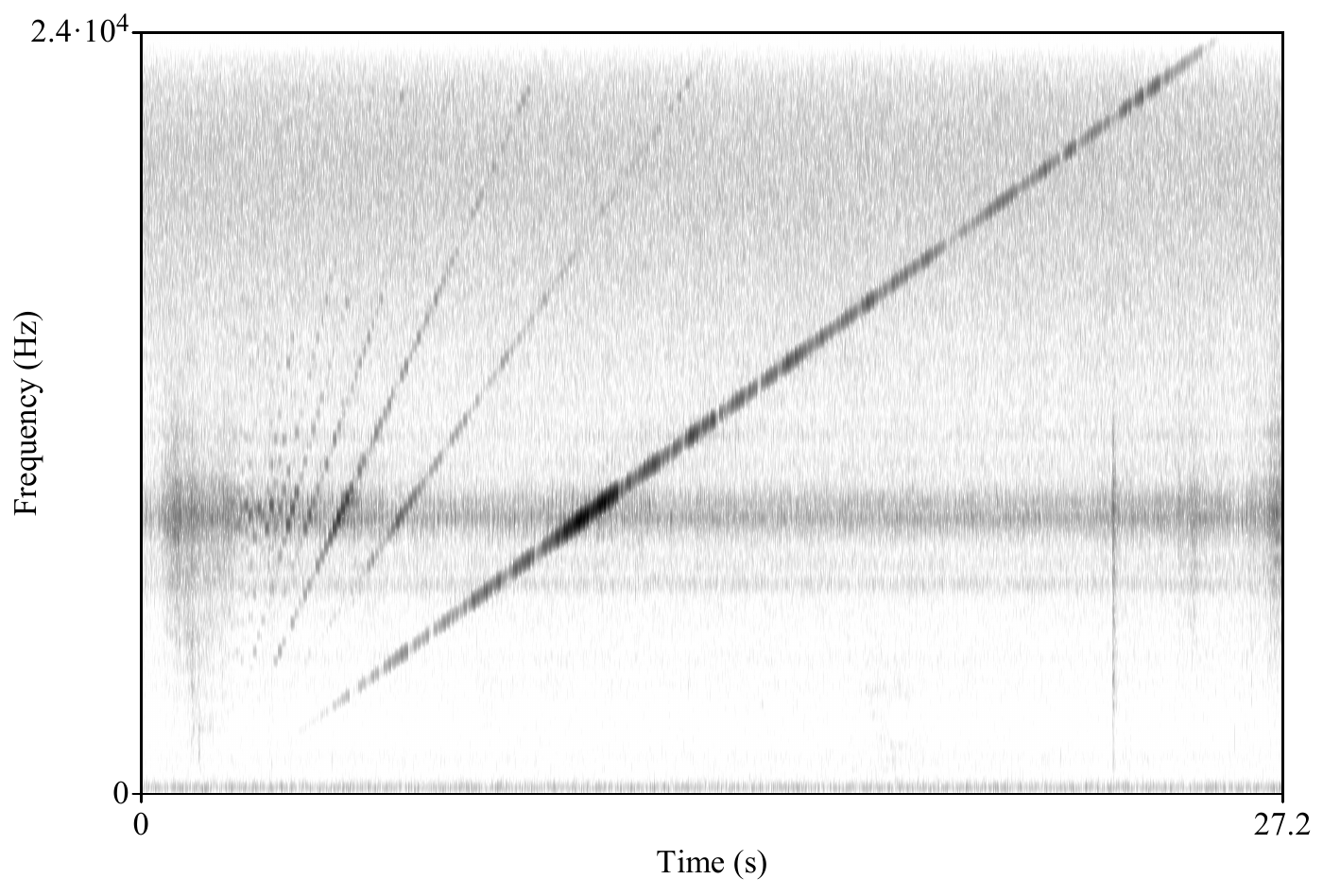}
		\caption{The sweep signal recorded from NUK (spectrogram view)}
	\label{fig:as2}
\end{figure}

\begin{figure}[]
	\centering
		\includegraphics[width=\linewidth]{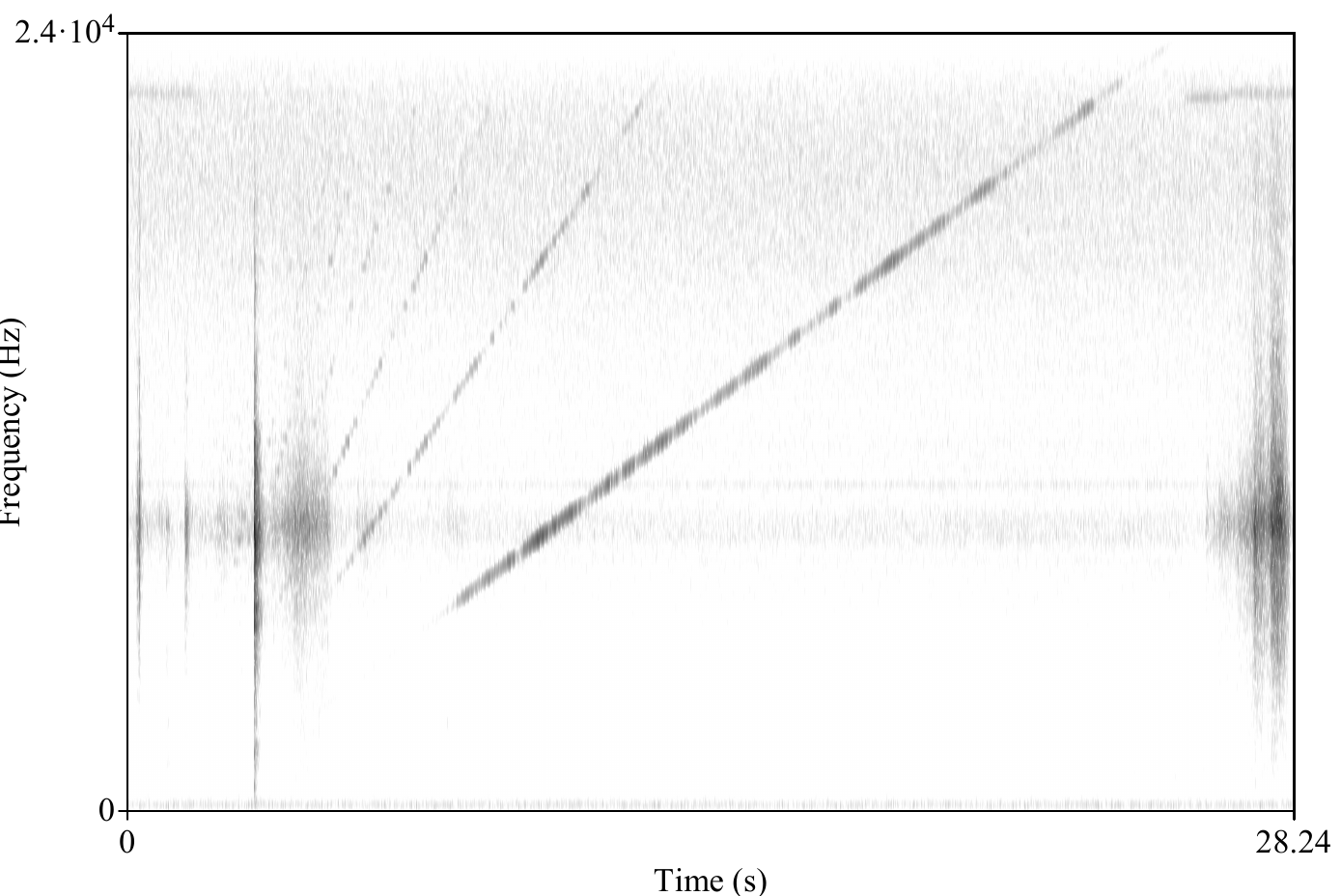}
		\caption{The sweep signal recorded from IOT (spectrogram view)}
	\label{fig:as3}
\end{figure}

We examined the frequency bands of each of the six PSUs by analyzing the \textit{chirp signal} they generate. This signal, also known as a \textit{sweep signal}, is generated by gradually increasing the switching frequencies of the PSU from the minimum to the maximum in a specific range. In our case, we examined the sweep signal in the band of 0-24kHz which can be received by a nearby smartphone. For the analysis of the chirp signal, the recording smartphone was located 20 cm away from the tested device. Table \ref{table:bands} summarizes the analysis of the sweep signals recorded from the six computers. It presents the frequency band generated from each computer and an estimation of the signal quality based on the SNR ranges measured from the entire sweep signal. 

\begin{table}[!h]
	\centering
	\caption{The frequencies of the audible bands of for six PSUs under test}
	\label{table:bands}
	\begin{tabular}{@{}lll@{}}
		\toprule
		\#     & Best audible band & Signal quality (SNR range) \\ \midrule
		PC-1   & 2300Hz - 22000Hz  & High (30-35dB)         \\
		PC-2   & 2200Hz - 22000Hz  & High (30-35dB)         \\
		PC-3   & 5491Hz - 6223Hz, 8300Hz - 9000HZ & Low (20-30dB)      \\
		Server & 8000Hz - 17000Hz  & Low (10-20dB)          \\
		NUK    & 1800Hz - 24000Hz  & High (30-40dB)         \\
		IOT    & 3100Hz - 24000Hz  & Intermediate (20-30dB)         \\ \bottomrule
	\end{tabular}
\end{table}

The results show that the quality of the acoustic signal may vary among different types of computers. PC-1 and PC-2 generated a detectable signal in a band of 2300Hz-22000Hz, while PC-3 could only generate a weak signal on the fragmented bands of approximately 5500Hz-6220Hz and 8300Hz-9000Hz. The server computer generated a weak signal in a band of 8000Hz-17000Hz. The NUK computer generated a strong signal in the entire band of 1800Hz-24000Hz, and the IOT generated a intermediate signal in a band of 3100Hz-24000Hz. It is important to note that the strength of the signals across the spectrum are not uniform; the signal may be strong in certain frequency bands and weak in others. 
Based on the results, we chose PC-1, NUK, and IOT for the reminder of the evaluation. These computers produce strong signals, and they represent the common cases of a PC workstation, small form factor computer, and a low-power IoT device. The spectrogram of their chirp signals can be seen in Figures \ref{fig:as1}, \ref{fig:as2}, and \ref{fig:as3}, respectively.

\subsection{Signal-to-Noise Ratio (SNR)}
\begin{figure}
	\centering
	\includegraphics[width=0.75\linewidth]{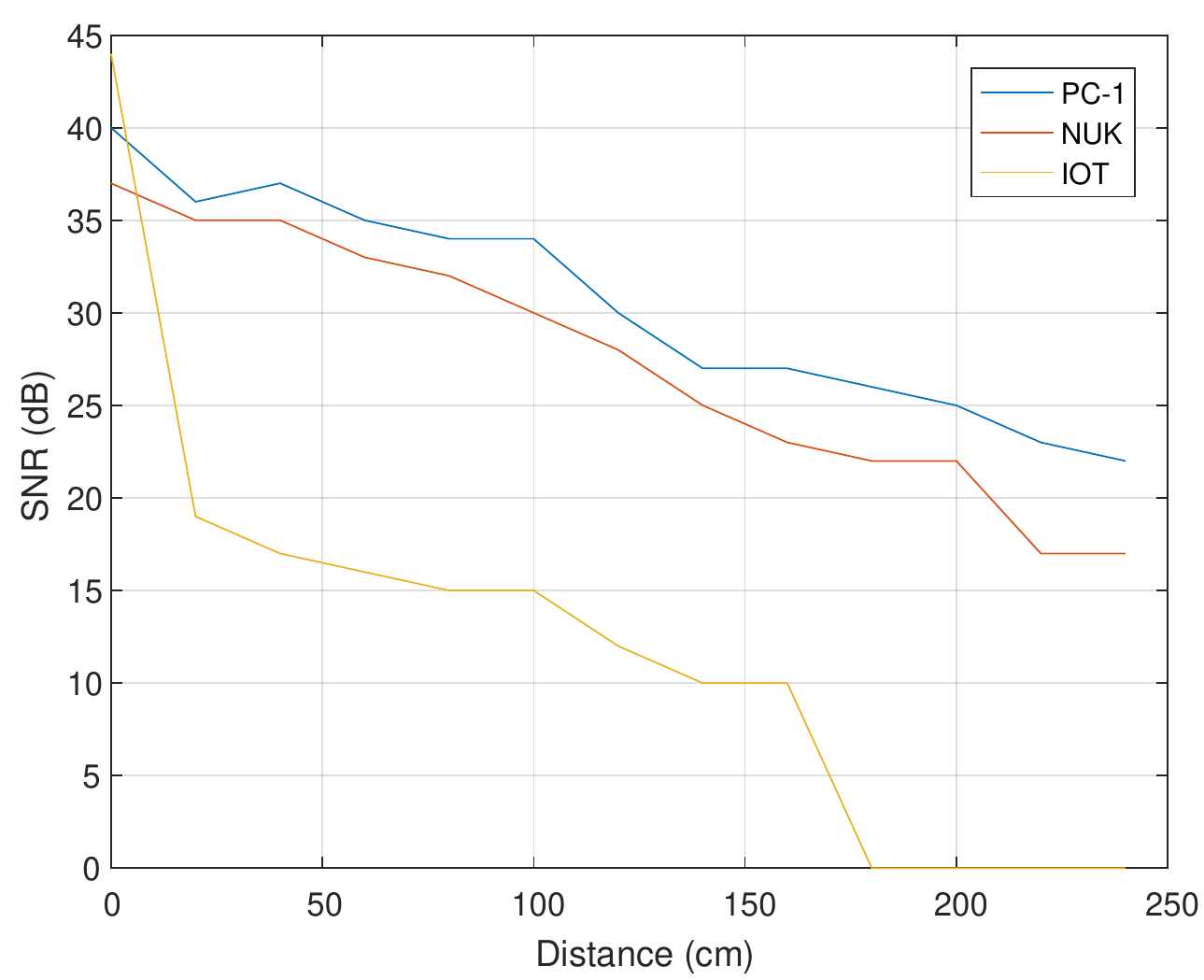}
	\caption{SNR levels measured at distances of 0-240cm}
	\label{fig:SNR}
\end{figure}

The acoustic signals generated from the PSU transformers and capacitors are limited in terms of strength. We measured the signal-to-noise ratio (SNR) to compare the level of a generated signal ($S$) to the level of the background noise ($N$) at different distances. This measure is correlated to the quality of the audible signal, e.g., an $S/N$ ratio greater than one indicates there is more signal than noise. We generated a signal for a time period of one second and compared it with the background noise. For the SNR measurements, we used the frequency band with the strongest signal for each PSU. The signals were recorded with the smartphone receiver located at distances of 0-250cm from the computer.  
The results are presented in Figure \ref{fig:SNR}.

As can be seen, the acoustic signal emanating from the IOT device could be received from a distance of about 150cm with an SNR of ~10dB. With PC-1 and NUK, the good quality signals could be received from a distance of 250cm with ~22dB and ~17dB, respectively.   

\subsection{Number of Cores}
\begin{figure}
	\centering
	\includegraphics[width=0.75\linewidth]{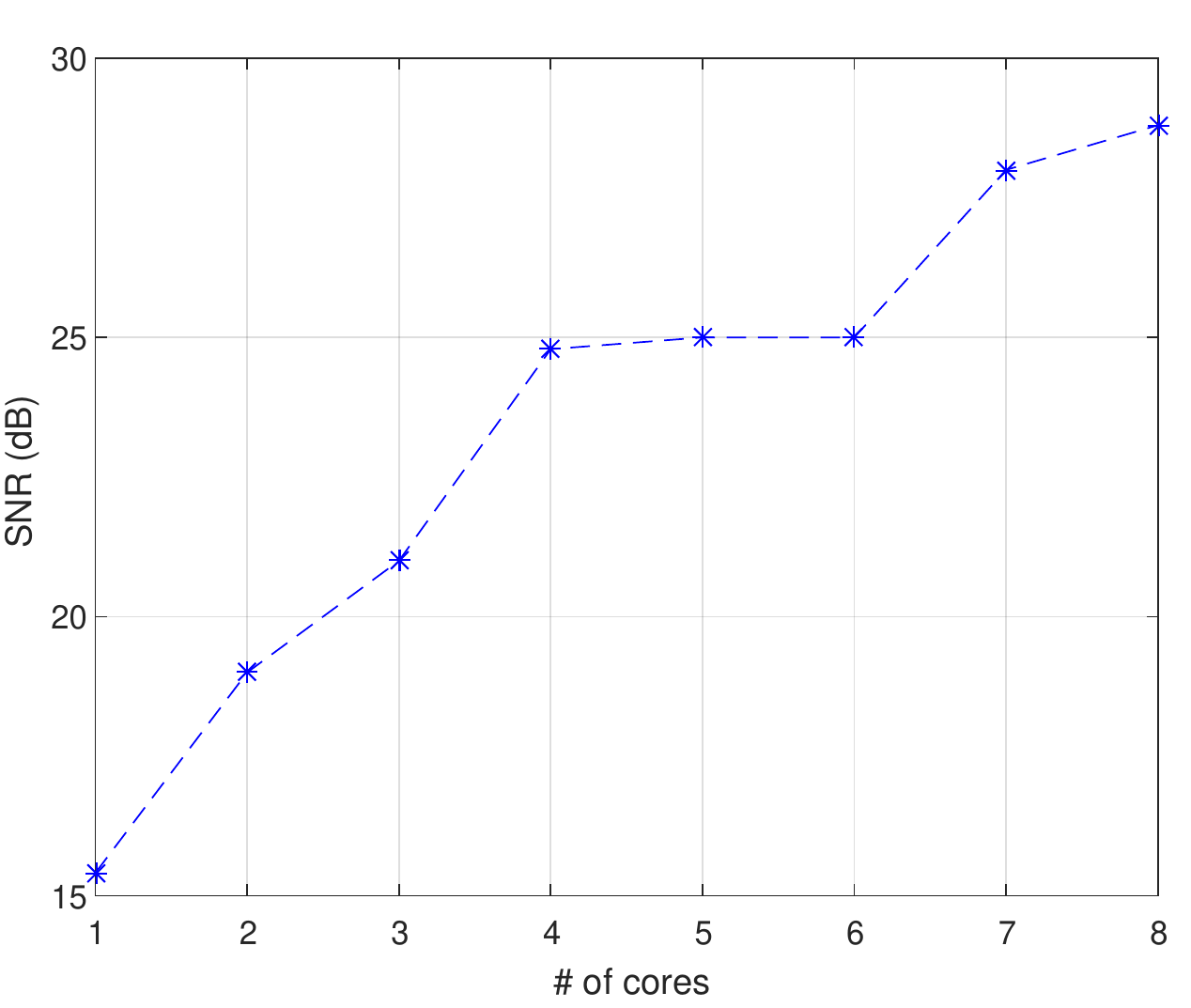}
	\caption{SNR levels measured with 1 to 8 cores}
	\label{fig:cores}
\end{figure}

As discussed, the number of cores used for signal generation directly influences the strength (amplitude) of the signal (i.e., more transmitting threads yield a stronger signal). Figure \ref{fig:cores} presents the SNR measurements of PC-1 with one, two, and four cores used for signal generation. In this test, we used one thread per core. The signal recorded at a distance of 20cm from the computers showed a gradual increase as the number of cores used for the signal generation increased (this ranged from an SNR of 15dB with one core to an SNR of 27dB with eight cores).  

\paragraph{Hyper-Threading} Note that modern Intel CPUs support hyper-threading technology \cite{marr2002hyper}. With this technology, each physical core exposes two logical (virtual) cores to the operating system. The CPU shares the workload between the logical cores when possible for better utilization. In the experiments, we bound the transmitting threads to the systems' logical cores rather than the physical cores, i.e., in a system with four physical cores and eight virtual cores we can potentially run eight concurrent transmitting threads.  

\subsection{Bit-Error-Rate (BER)}
As discussed in Section \ref{datatransfer}, the acoustic signal can be used for data transfer. We measured the bit error rate for PC-1, NUK, and IOT at distances of 0-240cm for  PC-1 and NUK and 100cm for the IOT device. For the tests, we used B-FSK modulation and the frequency band with the strongest signal for each PSU. 
We transmitted sequences of random bits, decoded them, and compared the results with the original data. The tests were repeated three times. The results for PC-1, NUK, and IOT are summarized in Tables \ref{BER-PC1}, \ref{BER-NUK}, and \ref{BER-IOT}, respectively, as described below.  


\begin{table*}[]
	\centering
	\caption{Bit error rates (BER) for PC-1}
	\label{BER-PC1}
	\begin{tabular}{lllllllllllllll}
		\hline
		\#   & bps & 0   & 20  & 40  & 60  & 80  & 100  & 120 & 140 & 160 & 180 & 200 & 220 & 240 \\ \hline
		PC-1 & 1   & 0\% & 0\% & 0\% & 0\% & 0\% & 0\%  & 0\% & 0\% & 0\% & 0\% & 0\% & 0\% & 5\% \\
		PC-1 & 10  & 0\% & 0\% & 0\% & 0\% & 0\% & 0\%  & 0\% & 0\% & 0\% & 0\% & 0\% & 0\% & 5\% \\
		PC-1 & 30  & 0\%   & 0\%   & 0\%   & 0\%   & 5\% & 10\% & -   & -   & -   & -   & -   & -   & -   \\ 
		PC-1 & 60  & 0\%   & 0\%   & 0\%   & 0\%   & - & - & -   & -   & -   & -   & -   & -   & -   \\ \hline
	\end{tabular}
\end{table*}

\begin{table*}[]
	\centering
	\caption{Bit error rates (BER) for NUK}
	\label{BER-NUK}
	\begin{tabular}{lllllllllllllll}
		\hline
		\#  & bps & 0   & 20  & 40  & 60  & 80  & 100  & 120 & 140 & 160 & 180 & 200 & 220 & 240 \\ \hline
		NUK & 1   & 0\% & 0\% & 0\% & 0\% & 0\% & 0\%  & 0\% & 0\% & 0\% & 0\% & 0\% & 5\% & 8\% \\
		NUK & 20  & 0\% & 0\% & 0\% & 0\% & 0\% & 0\%  & 0\% & 0\% & 0\% & 0\% & 0\% & 5\% & -   \\
		NUK & 50  & 0\%  & 0\%   & 0\%   & 0\%   & 5\% & 10\% & -   & -   & -   & -   & -   & -   & -   \\ 
		NUK & 60  & 0\%  &  0\% & -  & -   & - & - & -   & -   & -   & -   & -   & -   & -   \\ \hline
	\end{tabular}
\end{table*}

\begin{table}[]
	\centering
	\caption{Bit error rates (BER) for IOT}
	\label{BER-IOT}
	\begin{tabular}{llllllll}
		\hline
		\#  & bps & 0   & 20  & 40  & 60  & 80  & 100 \\ \hline
		IOT & 1   & 0\% & 0\% & 0\% & 0\% & 0\% & 2\% \\
		IOT & 20  & 0\% & 0\% & 0\% & 0\% & 0\% & 3\% \\
		IOT & 40  & 0\% & 0\% & 0\% & 0\% & 3\% & 3\% \\ 
		IOT & 80  & 0\% & 10\% & - & - & - & - \\ \hline
	\end{tabular}
\end{table}

\subsubsection{PC-1}
 With bit rWith bit rates of 1 bit/sec to 10 bit/sec we could maintain data transmission at distances of 240cm with a maximum BER of 5\%.ate of 30 bit/sec we could maintain data transmission at distances of 100cm with a maximum BER of 10\%. With bit rate of 60 bit/sec we could maintain data transmission at distances of 60cm with a maximum BER of 0\%. 
\subsubsection{NUK}
With bit rates of 1 bit/sec to 20 bit/sec we could maintain data transmission at distances of 220cm with a maximum BER of 5\% and distances of 240cm with a maximum BER of 8\%. With a bit rate of 50 bit/sec we could maintain data transmissions at distances of 100cm with a maximum BER of 10\%. With a bit rate of 60 bit/sec we could maintain data transmission at short distances of 20cm with a maximum BER of 0\%.
\subsubsection{IOT}
With bit rates of 1 bit/sec, 20 bit/sec, and 40 bit/sec, we could maintain data transmission at distances of 100cm with a maximum BER of 2-3\%. With a bit rate of 80 bit/sec, we could maintain data transmission at distances of 20cm with a maximum BER of 10\%.

%
%
%
%
\subsection{Greater Distance}
We found that some power supplies produce strong acoustic signals in certain frequency bands. 
We were be able to receive acoustic signals up to six meters away from PC-4. However, the quality of the signal significantly decreases with the distance. From 4m,5m and 6m away we were able to receive the signals at maximal frequencies of 14kHz, 12kHz and 9kHz, respectively. Table \ref{tab:BER4} shows the results of PC-4. With bit rates of 5 bit/sec we could maintain data transmission at distances of 6m away with a maximum BER of 1.2\%. 
 
\begin{table}[]
	\centering
	\caption{Bit error rates (BER) for PC-4 with 5 bps}
	\label{tab:BER4}
	\begin{tabular}{@{}lllllll@{}}
		\toprule
		\#   & 1m  & 2m  & 3m    & 4m    & 5m    & 6m    \\ \midrule
		PC-4 & 0\% & 0\% & 1.1\% & 1.2\% & 1.2\% & 1.2\% \\ \bottomrule
	\end{tabular}
\end{table}

\subsection{Threat Radius}
\begin{figure}
	\centering
	\includegraphics[width=0.7\linewidth]{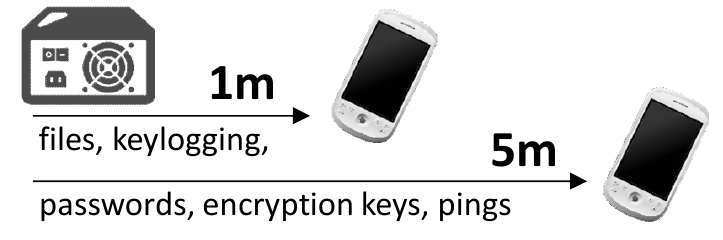}
	\caption{The distances and relevant data for exfiltration}
	\label{fig:distances}
\end{figure}

The above results indicates that the covert channel can be used to transfer data with bit rates of 0-40 bit/sec at short distances of 0-100cm. Such bit rates could be used to transmit binary data, keystrokes logging, text files, and so on. With slower bit rates of 1-10 bit/sec, the covert channel  is effective for even greater distances (200-300cm with PC-1 and NUK, 200cm-600cm with PC-4). The slower bit rates could be used to transfer a small amount of data, such as  short texts, encryption keys, passwords, and keystroke logging, as seen in Figure \ref{fig:distances}.  

\subsection{Stealth}
The transmitting code shown above requires no elevated privileges and can be initiated from an ordinary user space process. The code consists of basic CPU operations such as busy loops, which do not expose malicious behavior, making it highly evasive an unable to detect by static and dynamic (runtime) malware detection solutions.

\section{Countermeasures}
\label{sec:counter}
There are four main categories of countermeasures that can be used to defend against the proposed covert channel: zones separation, signal detection, signal jamming, and signal blocking. 
\paragraph{Zoning}
In procedural countermeasures the 'zoning' approach may be used. In this approach sensitive computers are kept in restricted areas in which mobile phones, microphones, and electronic equipment are banned. The zoning approach (also referred to as 'red/black separation') is discussed in \cite{NSTISSAM75:online} as a means of handling various types of acoustic, electromagnetic, and optical threats. However, zoning is not always possible, due to practical limitations such as space and cost. In our case, recording devices such as mobile phones should be banned from the area of air-gapped systems or be kept at a certain distance from them. 

\paragraph{Signal Detection} 
Host based intrusion detection systems (HIDS) may continuously trace the activities of running processes in order to detect suspicious behavior; in our case, a group of threads that abnormally regulates the switching frequency would be reported and inspected. Such a detection approach would likely suffer from  false alarms, since many legitimate processes use CPU intensive calculations that affect the processor's workload \cite{carrara2016survey}. Another problem in the runtime detection approach is that the signal generation algorithm (presented in Section \ref{sec:trans}) involves only non-privileged CPU instructions (e.g., busy loops). Monitoring non-privileged CPU instructions at runtime necessitates that entering the monitored processes enter a \textit{step-by-step} mode, which severely degrades performance \cite{guri2015gsmem}. Software based detection also suffers from an inherent weakness in that it can easily be bypassed by malware using evasion techniques \cite{cardenas2011attacks}. In our case, the malware may inject the transmitting threads into a legitimate, trusted process to bypass the security mechanisms. 

Hardware-based countermeasures may include noise detector devices which monitor the spectrum at a range of frequencies. Such products exist \cite{Products93:online} but are prone to false alarms due to natural environmental noises \cite{carrara2016survey}. Jamming the PSU signal by the generating background noise at a specific ranges \cite{AudioJam82:online} is also possible but not applicable in some environments, particularly in quiet settings. In our case, the frequency band is 0-24kHz which is include the audible region.

\paragraph{Signal Jamming} 
Jamming the whole frequency band will generate a noticeable amount of environmental noise which may disturb users. Physical isolation in which the computer chassis is built with special noise blocking cover is also an option, but it is costly and impractical on a large-scale. Carrara \cite{carrara2014acoustic} suggested monitoring the audio channel for abnormally energy peaks, in order to detect hidden transmissions in the area. In our case, the ultrasonic frequency range above 18kHz should be scanned (continuously) and analyzed. However, as noted in \cite{carrara2014acoustic}, if the monitoring devuce is far from the transmitter this approach may not be effective.

\paragraph{Signal Limiting/Blocking} 
Although there are quiet PSUs that limit the acoustic noise emitted from their internal components, this feature does not \textit{hermetically} prevent the emission of noise \cite{handbookbillings2011switchmode}. Physical isolation in which the computer chassis is enclosed within a special noise blocking cover is also an option, but it is costly and impractical on a large scale.

\section{Conclusion}
\label{sec:conclusion}
In this paper, we show that malware running on a computer can use the power supply as an out-of-band speaker. A code executed in the system can intentionally regulate the internal \textit{switching frequency} of the power supply, and hence control the waveform generated from its capacitors and transformers. This technique allows sonic and ultrasonic audio tones to be generated from a various types of computers and devices even when audio hardware is blocked, disabled, or not present. We show that the POWER-SUPPLaY code can operate from an ordinary user-mode process and doesn't need hardware access or root-privileges. This proposed method doesn't invoke special system calls or access hardware resources, and hence is highly evasive. We present the implementation details and evaluation, and provide the measurement results. By using POWER-SUPPLaY, we could acoustically exfiltrated data from audio-less systems to a nearby mobile phone at a distance of 2.5 meters with a maximal bit rate of 50 bit/sec. 

\balance
\bibliographystyle{ieeetran}
\bibliography{../../../AirGap}

\end{document}